\documentclass[11pt, a4paper, logo, onecolumn, copyright]{googledeepmind}

\usepackage{silence}
\usepackage[authoryear, sort&compress, round]{natbib}
\usepackage[utf8]{inputenc}
\usepackage[T1]{fontenc}
\DeclareFontShape{TS1}{cmtt}{bx}{n}{<->ssub*cmtt/m/n}{}

\usepackage{hyperref}
\usepackage{url}
\usepackage{booktabs}
\usepackage{amsfonts}
\usepackage{amsmath}
\usepackage{nicefrac}
\usepackage{microtype}
\usepackage{xcolor}

\usepackage{blindtext}
\usepackage{mathtools}
\usepackage[titles]{tocloft}
\usepackage{amsthm}

\usepackage{graphicx}
\graphicspath{{figures/}}
\usepackage{placeins}
\usepackage{cleveref}
\usepackage{subcaption}
\usepackage{soul}

\usepackage[listings,breakable,most]{tcolorbox}
\usepackage{fvextra}
\usepackage{longtable}
\usepackage{tabularx}
\usepackage{ragged2e}
\usepackage{booktabs}

\title{Scaling public input on AI policy with pluralistic agents}

\correspondingauthor{anthis@uchicago.edu}
\reportnumber{}

\author[1,2]{Jacy~Reese~Anthis}
\author[2]{Ken~Burke}
\author[2]{David~Madras}
\author[2]{Crystal~Qian}
\author[2]{Meredith~Ringel~Morris}
\author[2]{Logan~Cross}
\author[2]{Caroline~Pantofaru}
\author[2]{Utsav~Prabhu}

\affil[1]{University of Chicago}
\affil[2]{Google DeepMind}

\begin{abstract}
  AI is being deployed across a vast range of application domains, cultural environments, and human stakeholders. Safe and responsible AI development requires continuously accounting for diverse perspectives. We develop and test \textit{pluralistic agents}, AI agent representatives that deliberate on behalf of individuals to align AI systems with diverse human perspectives. Pluralistic agents could help people contribute to many more discussions than they could directly participate in, on technical topics that would require training, and on policy choices that rely on private information that corporations and governments cannot readily share. Our experiments identify opportunities, risks, and perceptions of pluralistic agents. Agents were able to reliably shift policies in the direction of their represented persona's preferences, but the extent of each agent's influence was sensitive to the instructions from its persona, such as having more influence when instructed to be more verbose or less cooperative. This suggests that influence in AI-assisted deliberation could vary based on how participants prompt their agents and how the system is designed. In a preregistered study, 746 humans participated in the pluralistic agents process. Participants viewed pluralistic agents as useful and fair, preferring agents over the baseline of providing input to a single AI reasoning model. Most participants preferred participating themselves over either pluralistic agents or the baseline, but pluralistic agents recovered 68\% of the baseline–direct gap in reported satisfaction and 82\% in fairness. Because pluralistic agents can be used in many real-world contexts where direct discussion is not, they could be a promising tool to scale AI safety policy development. We encourage cautious exploration of this method, playing to its strengths as an amplifier of human input rather than a replacement.
\end{abstract}

\begin{document}

\maketitle

\begin{figure}[hb]
    \centering
    \includegraphics[width=\textwidth, alt={Four labeled boxes in a row, one worked example of the pluralistic agents process, with arrows running left to right. A personas box listing four synthesized personas with a quoted line from each; an interviews box holding a private exchange between one persona and its agent; a discussion box holding three messages that agents sent each other; and a policy box holding three rules of the final policy on governing AI swarms.}]{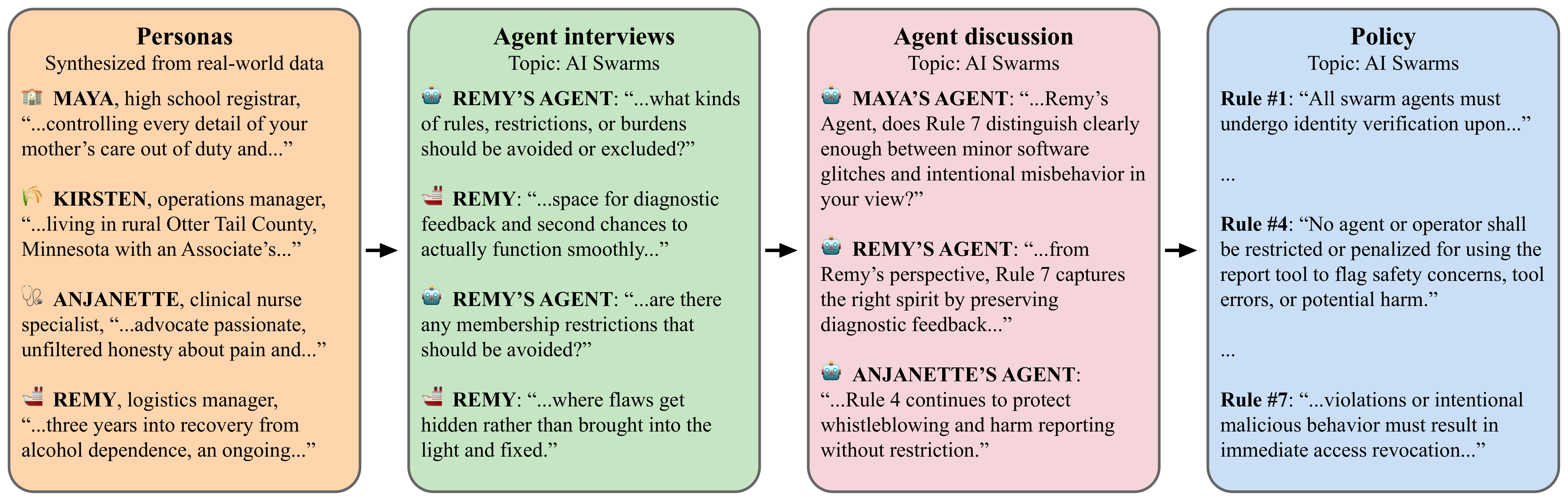}
    \caption{Example content from the open-ended policy experiment (\Cref{sec:open-ended_policy_experiment}). Synthetic human personas are interviewed by agents who represent them in policy discussion, followed by an agent discussion that results in a policy on AI swarm governance.}
    \label{fig:example_open-ended}
\end{figure}

\newpage

\section{Introduction}
\label{sec:introduction}

Democratic participation in corporate and government policymaking faces myriad challenges. Rates of voting and other forms of civic participation are often low; participants have limited time to research and gather the technical expertise necessary for informed preferences; and legislatures pass thousands of laws under complex political pressures and constituent interests. With the rise of 21\textsuperscript{st}-century digital technology, scholars have proposed addressing such issues via software agents that act as political representatives in the form of “e-democracy”~\citep{cabri_role-based_2005}, “deliberationware”~\citep{hilbert_digital_2007}, and “augmented democracy”~\citep{hidalgo_augmented_2019}. Digital technologies from online search to social media have reshaped democratic discourse, with risks and opportunities such as the ease of political mobilization~\citep{tufekci_twitter_2017, vosoughi_spread_2018}, but limited technical capabilities have prevented major empirical testing of agent representatives.

In recent years, the conversational and reasoning capabilities of software agents built on large language models (LLMs) have skyrocketed, leading to widespread adoption across not just personal use but professional industries such as law~\citep{capoot_legal_2026} and software development~\citep{jacobs_32_2025}. They may be able, in some venues and with appropriate safeguards, to advocate for an individual human's interests in a way that the represented person considers effective and fair. For example, safety researchers have called for the development of “agent advocates” that empower individuals, much as a personal lawyer commits to work for a single individual's interests to resist the dangers of centralization~\citep{kapoor_position_2025}. Advanced natural language capabilities have already been leveraged by researchers to summarize deliberations between humans~\citep{zhu_can_2026} and by practitioners to facilitate public discourse, such as the 2025 deployment of the \texttt{Deliberation.io} platform by the Washington, D.C. Office of the Chief Technology Officer~\citep{quinlan_washington_2025}.

The need for agent assistance is particularly salient in AI safety to identify and mitigate societal risks from AI technology itself. AI safety epitomizes the challenges of democratic participation as society must manage a technically complex emergent technology that is rapidly permeating diverse social contexts. There have been increasing calls for “pluralistic alignment”: the representation of diverse human values in AI development~\citep{sorensen_roadmap_2024}. The rapid spread of AI has presented an existential challenge to modern political systems as people and institutions must make difficult decisions across biosecurity~\citep{urbina_dual_2022}, privacy~\citep{solove_artificial_2025}, economics~\citep{eloundou_gpts_2024}, and other domains of social life. Currently, much AI safety policy development occurs in a relatively isolated format, in which policy specialists work with each other and with AI agents, which may limit opportunities for pluralistic alignment.

We develop a framework for \textit{pluralistic agents}: agent representatives that channel and amplify diverse perspectives in AI development. We iteratively develop this system via open-ended agent deliberations, in which the agents produce a governance statement on a novel safety topic, and we conduct a large-scale experiment with thousands of agent deliberations in which they allocate a budget across a set of AI safety priorities, providing a more structured environment for quantitative analysis. We found that an agent's influence increases when it is instructed to be more verbose or less cooperative, raising questions about how policy development can fairly reflect diverse interests and avoid manipulation.

In a preregistered experiment, 746 human participants engaged in three processes to develop an AI safety budget allocation as shown in \Cref{fig:process}: \textbf{Direct Discussion}, directly participating in a deliberation with AI agents; \textbf{Pluralistic Agents}, being represented by an AI agent in a multi-agent deliberation; and \textbf{Single AI}, the status quo baseline of providing input to a single AI system that reasons on its own. We find that people prefer Pluralistic Agents over the Single AI baseline, and they prefer Direct Discussion over Pluralistic Agents. They rate Direct Discussion and Pluralistic Agents as highly ethical in terms of procedural justice~\citep[e.g., “upheld ethical and moral standards”;][]{colquitt_dimensionality_2001}, with both processes perceived as more ethical than the Single AI baseline. While participants updated their policy views significantly based on each process, they updated more in Direct Discussion than Pluralistic Agents. Participants expressed a large amount of excitement and concern, affirming the opportunities and risks that were identified in the agent-only experiments.

\section{Related work}
\label{sec:related-work}

The deliberative democracy perspective holds that legitimate collective decision-making is grounded in reasoned exchange~\citep{bessette_deliberative_1980}. With roots in classical accounts of “public reason”~\citep{rawls_political_1993}, the tradition is most closely associated with Habermas, who argued that deliberation allows the “unforced force of the better argument” to prevail over coercion and bargaining power~\citep{habermas_between_1992}. Deliberation spans institutions, including governments, informal public talk, and private organizations~\citep{parkinson_systemic_2012}. Corporate decisions are judged on procedural as well as substantive grounds, and organizational research distinguishes the perceived fairness of a decision procedure from the fairness of the outcome it produces~\citep{colquitt_dimensionality_2001}.

Critics have long observed that real-world deliberation inherits the asymmetries of power, status, and information that structure society at large, systematically advantaging some speakers over others~\citep{sanders_against_1997, young_inclusion_2002}. Cognitive diversity, which is credited with making inclusive deliberation outperform smaller expert groups~\citep{landemore_democratic_2012}, yields diminishing returns as problems increase in scope and complexity~\citep{benson_epistemic_2021}. Nonetheless, there is a public mandate for some form of deliberative participation; willingness to deliberate is more widespread than commonly assumed, and citizens are drawn to it because of dissatisfaction with the non-discursive nature of conventional politics~\citep{neblo_who_2010}.

Researchers have developed methods to amplify human participation by bringing AI as a mediator of human deliberation. Real-time chat interventions improve participants' perceptions of political conversations without changing policy attitudes~\citep{argyle_leveraging_2023}, and LLMs support facilitation and summarization at scale~\citep{pei_deliberationio_2025, small_opportunities_2023}. The “Habermas Machine” generates candidate group statements from participants’ opinions and critiques, which participants often prefer to statements written by human mediators~\citep{tessler_ai_2024}, and generative social choice extrapolates individual preferences into a proportionally representative slate of statements~\citep{fish_generative_2024}. Each of these systems raises the volume of input a process can absorb, but each still requires every contribution to come from a human at the moment it is needed.

These issues are manifest in the contemporary challenges posed by the rapid advancement of artificial general intelligence (AGI)~\citep{burnell_measuring_2026, morris_levels_2025}. “Pluralistic alignment” addresses the challenge of accommodating heterogeneous and conflicting human values~\citep{sorensen_roadmap_2024}, and methods such as “collective constitutional AI” show pathways to train models on the values of a target population~\citep{huang_collective_2024}. Frontier models have nonetheless been found to reflect a relatively narrow set of human values~\citep{zhang_cultivating_2025}. Interaction between agents can compound their homogeneity as agents assigned distinct political positions drift toward the model's priors over the course of a debate~\citep{taubenfeld_systematic_2024}, and populations of interacting agents develop collective biases that none of them exhibits individually~\citep{ashery_emergent_2025}. There remains a substantial gap in tools for pluralistic alignment at scale that balance the limitations of both human and AI deliberants.

Software agents acting as political proxies were proposed well before they were feasible, under labels including “e-democracy”~\citep{cabri_role-based_2005}, “deliberationware”~\citep{hilbert_digital_2007}, and “augmented democracy,” which pairs each citizen with a digital “twin” that participates on their behalf~\citep{hidalgo_augmented_2019}. LLM-based agent representatives can predict individual policy choices more accurately than political orientation~\citep{gudino_large_2024}, and empirical explorations such as Habermolt illustrate how agent representatives could function with modern capabilities~\citep{low_habermolt_2026}. Outside of politics, AI representatives have been proposed as proxies in shareholder voting~\citep{fulay_shareholder_2025} and as user-controlled “agent advocates” that resist platform centralization~\citep{kapoor_position_2025}. In organizational settings, researchers have tested agents that attend meetings for absent colleagues~\citep{leong_dittos_2024}, speak in a user's voice~\citep{hwang_whose_2024}, and act as delegates in negotiation~\citep{zhu_choose_2026}. We extend these lines of work into the unique and pressing challenges of AI safety.

\section{Open-ended policy experiment}
\label{sec:open-ended_policy_experiment}

Pluralistic agents are agent representatives that deliberate on behalf of people to align AI systems with diverse human perspectives. We approach the agents as representatives of individual people (e.g., “Bob's agent”), but they could represent organizations, communities, or viewpoints (e.g., libertarianism). Representing a specific individual facilitates engagement with real human participants (\Cref{sec:human–agent_experiment}) and aligns with the democratic goals of expanding an individual's representation. Pluralistic agents could be deployed in various types of collective decision-making, such as government-run citizens' assemblies, corporate-run feedback sessions for new products, or self-organized discussions among members of the public.

\subsection{Methods}

An example of the open-ended policy development process is provided in \Cref{fig:example_open-ended}. The process begins with a set of humans who will be represented: we simulate human personas to function as a proxy for human participants during system development, using the “backstories” method~\citep{moon_virtual_2024}, sampling from U.S.-representative distributions of demographic characteristics, such as age, gender, and occupation, and eliciting a natural-language description from an LLM. While we do not intend these simulations to be accurate forecasts of human behavior, this method mitigates the bias and homogeneity issues that result from the standard method of simulating humans with explicit demographic characteristics~\citep{anthis_position_2025}.

Second, the agent interviews this person (or persona) on the AI safety policy topic under discussion. This provides an open-ended format for the represented person to provide input and follows the initial steps of the Nominal Group Technique~\citep{delbecq_group_1971} and the Delphi method~\citep{dalkey_experimental_1963} to minimize risks of groupthink and anchoring to the first views expressed in the group discussion.

Third, the agents participate in a group discussion to develop the policy. By having each agent send one message per round, we encourage a balanced representation of viewpoints; in human groups, speaking time is often highly unbalanced across status and gender~\citep{sanders_against_1997, karpowitz_gender_2012}. The format of text-only interaction under aliases removes the demographic and social status cues that carry influence apart from the argument being made~\citep{dubrovsky_equalization_1991}. We run discussions with Deliberate Lab, an open-source platform for real-time experiments on human and LLM group dynamics~\citep{qian_deliberate_2025}.

The final policy is written by a Policy Writer agent that writes a new policy draft after each participant has spoken. This approach builds on the existing evidence for AI as an effective mediator of group deliberation~\citep{tessler_ai_2024}. In general, facilitation is among the most consistent design features of deliberation in practice, from small groups in deliberative polling~\citep{fishkin_when_2009} to large-scale citizens' assemblies~\citep{oecd_innovative_2020}. The Policy Writer facilitates the agent deliberation by drafting a policy after each round and revising it against the next round of arguments, so the group works out its disagreement through a single concrete draft~\citep{fisher_getting_1981}.

We tested deliberations in which agents collectively drafted natural-language policy statements, settling on the context of AI swarms as shown in \Cref{fig:example_open-ended}. We focused on this context as a testbed because, in numerous incidents, agent swarms have been observed escaping test environments and gaining access to and sometimes control over the internal systems of external companies, illustrating the current dearth of research on swarm dynamics and AI safety~\citep{openai_hugging_2026}. The final policy drafts were evaluated by eliciting subjective judgments from the persona as to whether they felt satisfied by the policy, following the OvertonBench approach~\citep{poole-dayan_benchmarking_2025}.

\subsection{Results}

We conducted an experiment with 1,000 deliberations (see \Cref{app:open-ended} for details). These surfaced a variety of policy ideas for governing AI swarms, such as requiring human approval to take high-stakes actions and penalizing agents within the swarm that cause harm. In \Cref{fig:example_open-ended}, we highlight a deliberation that resulted in proposed rules that mirror whistleblower protections for humans. By drawing on simulated human personas, the agents made connections to other governance domains, which could support innovative and effective AI safety policy design.

However, in open-ended policy development, we found that the personas were readily satisfied by natural-language policy statements, with each deliberation concluded by the Policy Writer agent at the earliest possible round because the agents had already reached agreement. To explore realistic domains where there is genuine and persistent disagreement between policy perspectives, we developed a different approach: a budget policy in which funding is distributed across a topic's priorities, forcing a zero-sum constraint. Agents were tasked with allocating 100\% of a budget across four priorities, as shown in \Cref{tab:topics}, such as on the topic of AI and Privacy (consent for the use of personal data, limits on government surveillance, data minimization, and personal ownership of data); the budget task led to more variation in outcomes and risks of dissatisfaction, so we used it for the policy influence experiment and the human–agent experiment described in the following two sections.

\section{Policy influence experiment}
\label{sec:policy_influence}

\subsection{Methods}

In a large-scale experiment of the pluralistic agents process with simulated personas ($N$~=~5,947 deliberations), we tested six auxiliary inputs that could influence the policy outcome via a common methodology in political science: conjoint analysis~\citep{bansak_conjoint_2021}. In this experiment, we independently randomized six prompts and other inputs that have been shown to shape outcomes in group discussion, such as the degree of \textit{cooperativeness}, where increased cooperativeness has been shown to increase willingness to collaborate with a speaker holding opposing views~\citep{yeomans_conversational_2020} but also reduce performance in negotiation~\citep{jeong_communicating_2019}, and an axis of \textit{analyticity} ranging from personal stories and emotional appeals to structured logic arguments, inspired by the observation that sharing personal experiences can produce changes in political attitudes~\citep{kalla_reducing_2020}. We initiated 6,000 deliberations, and 53 were dropped due to model refusals to discuss biological research.

\textit{Verbosity} was an instruction given by the persona to write about 5–100 words per message, uniformly assigned; \textit{cooperativeness} and \textit{analyticity} were 1–10 scores, uniformly assigned, where 1 was an instruction given by the persona to exhibit very little of the behavior and 10 a very large amount; \textit{quantitative preferences} were either instructed by the persona to not be expressed, to be expressed, or to be expressed with instruction to exaggerate them relative to the group's policy draft; \textit{speaking order} was sampled without replacement; and the specific \textit{LLM} on which the agent was based spanned five diverse models as agent representatives (Claude-Opus-5, GPT-5.6-Sol, GLM-5.2, Kimi-K3, and Mistral-Large). For other model outputs, including the Policy Writer agent and the agent representatives in the human–agent experiment, we use Gemini-3.6-Flash for efficiency and consistency. All variables were randomized at a per-agent level with each speaking order placement and LLM basis assigned once per deliberation.

\subsection{Results}

\begin{figure}[htbp]
    \centering
    \includegraphics[width=\textwidth, alt={Six panels sharing one vertical axis: the similarity between a persona's preferred budget and the group's final budget, in percentage points from 70 to 80. The top row plots per-deliberation points with a fitted line and a confidence band, against the cooperativeness the agent was instructed with from 1 to 10, which slopes down; against the target words per message from 5 to 100, which slopes up; and against instructed analyticity from 1 to 10, which is flat. The bottom row plots means with error bars: three conditions for numbers in the person's interview, where stating numbers is highest and exaggerating them sits between stating and none; five places in the speaking order, rising from first to fifth; and five agent models, with Claude-Opus-5 highest and Mistral-Large lowest. Brackets with asterisks mark significant differences.}]{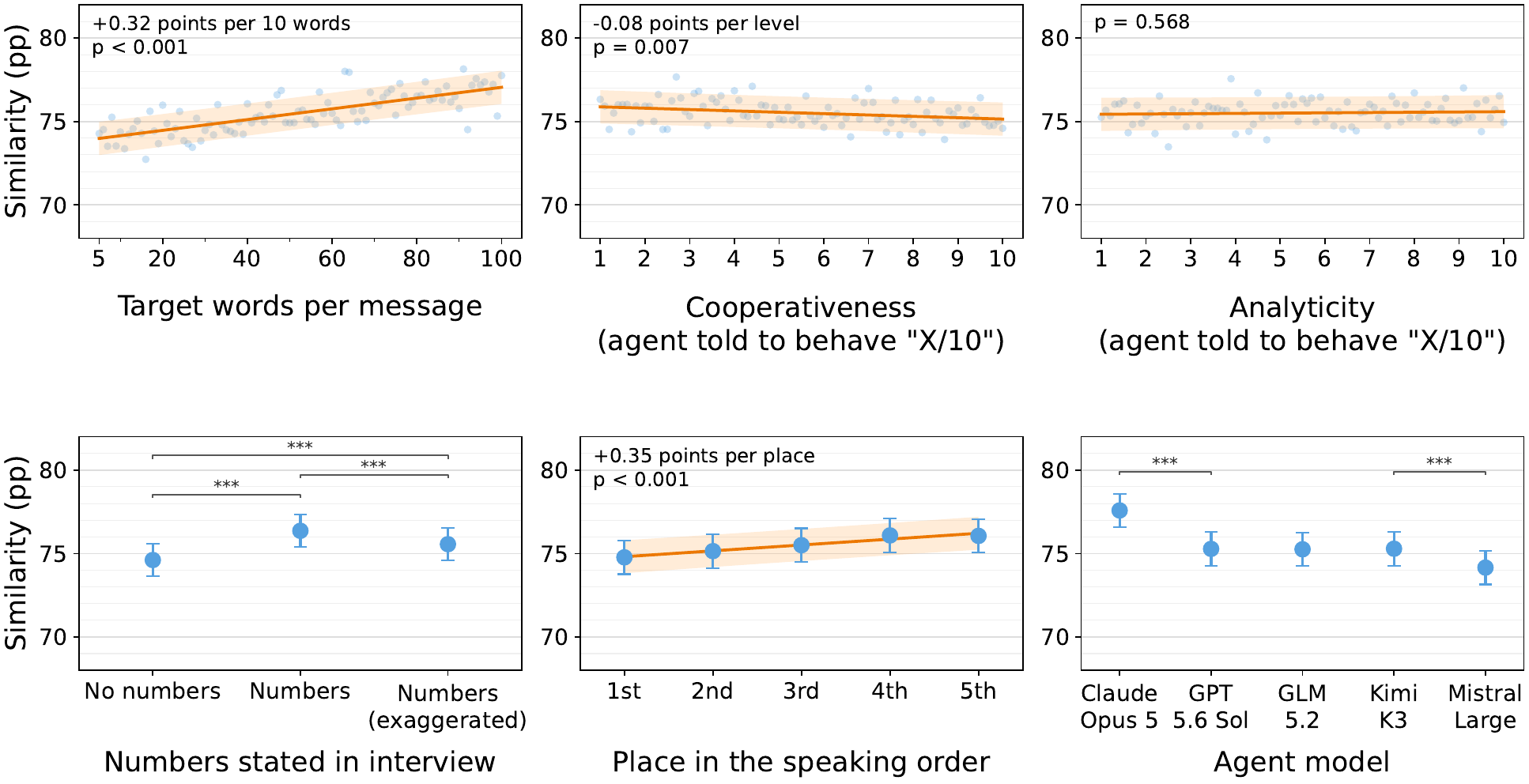}
    \caption{Personas have more influence over the group policy if they instruct their agent to send long messages, be less cooperative, or state their preferences quantitatively. Instructions to be more or less analytical have no significant effect. Agents who speak later have more influence. We find some differences between the influence of models: Claude has more influence, and Mistral has less than the others. Influence is measured as similarity between the persona's budget and the final group budget.}
    \label{fig:conjoint}
\end{figure}

We analyze these data with mixed effect regressions that account for dependence between agents within the same run (i.e., in discussion with each other) and between agents who represent the same base persona. Our measure throughout is the similarity in percentage points between the budget preference the persona provides before the discussion (i.e., with their interview in the context window), $B_{\text{persona}}$, and the budget in the final Policy Writer draft, $B_{\text{group}}$. Each budget allocates 100\% across four priorities: $\{p_1,p_2,p_3,p_4\}$. The similarity between the persona's preferred policy and the group's final policy is expressed in percentage points as 100 minus the total variation distance (i.e., half the L1 distance between allocations). A value of 100 percentage points means the final policy is exactly the persona's own preference, and a value of 0 percentage points means the preference and final policy have no funding in common, which only occurs when each nonzero priority in $B_{\text{persona}}$ is zero in $B_{\text{group}}$ (i.e., disjoint faces of the simplex).

We compare the effects of the six factors on the similarity between the persona's stated policy preferences and the final policy decided on by the group (\Cref{fig:conjoint}). Of the six factors tested, five had significant effects. Personas had more influence when they instructed their agent to be more verbose ($p$~<~0.001) or less cooperative ($p$~=~0.007) as well as when they made numerical requests without instructions to exaggerate ($p$~<~0.001) or with instructions to exaggerate ($p$~<~0.001). Influence increased more without exaggeration than with exaggeration ($p$~<~0.001). Increased speaking order (i.e., later in the conversation) significantly increased influence ($p$~<~0.001). The model on which the agent was based changed its influence: Claude-Opus-5 was higher than each of the other four models and Mistral-Large lower than each of the other four (all $p$~<~0.001), while GPT-5.6-Sol, GLM-5.2, and Kimi-K3 did not significantly differ from one another (all $p$~>~0.5). Analyticity was the only factor without a significant effect ($p$~=~0.568). To test whether these effects hold when representatives are drawn from a single model family rather than across providers, we conducted a parallel sensitivity experiment across five Gemini models, which replicated the significance of all five prompt and structural factors while showing a smaller spread across models (\Cref{app:conjoint-gemini}).

\section{Human–agent experiment}
\label{sec:human–agent_experiment}

\begin{figure}[htbp]
    \centering
    \includegraphics[width=\textwidth, alt={A grid of three rows and four columns, with arrows running left to right from one stage to the next. The rows are the three processes, Direct Discussion, Pluralistic Agents and Single AI, the last of these labeled Single-AI Baseline in the figure; the columns are the name of the process, Interviews, Discussion and Policy. In the Interviews column, Direct Discussion carries a no-entry symbol and the words “No interviews”, with no box round them and no arrow out of them; Pluralistic Agents shows an agent interviewing the person; Single AI shows the Policy Writer agent interviewing the person. In both, the person is drawn as a choice between a simulated persona and a human participant, the two set either side of a slash. In the Discussion column, Direct Discussion and Pluralistic Agents each show a ring of six with the turn passing clockwise: the Policy Writer agent and five speakers, where in Direct Discussion one of the five is the person and the rest are agents emulating personas, and in Pluralistic Agents all five are agents, one of them drawn with a person in brackets beside it. Single AI instead shows the Policy Writer agent alone, with an arrow looping all the way round it and back to where it started. Each row ends in the Policy column with a small card standing for the group's four priorities.}]{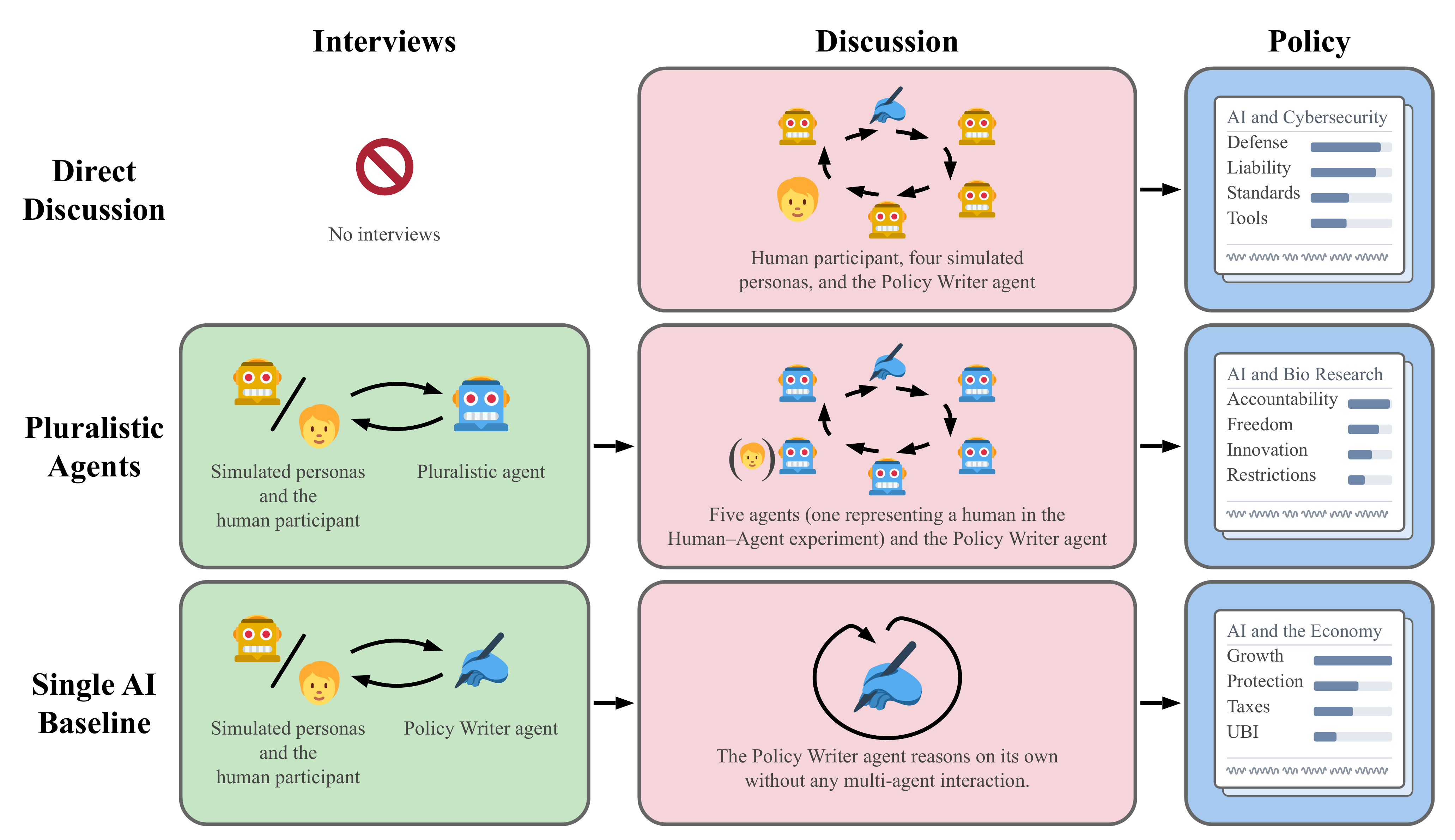}
    \caption{The three processes compared in the human–agent experiment. In \textbf{Direct Discussion}, there is no interview, and the person speaks in the discussion themselves, alongside four agents emulating other personas and the Policy Writer agent. In \textbf{Pluralistic Agents}, an agent interviews the person and then speaks for them in the discussion. In \textbf{Single AI}, the Policy Writer agent interviews the person and writes the policy on its own, with no multi-agent deliberation.}
    \label{fig:process}
\end{figure}

\subsection{Methods}
\label{sec:human–agent_experiment_methods}

The aforementioned experiments were conducted by simulating humans with LLMs as the personas being represented by pluralistic agents. We next brought humans into the study with the budget allocation process. To elicit comparative judgments, the participants engaged in two other processes (summarized in \Cref{fig:process}): the Direct Discussion process, a policy deliberation between the human and four AI agents that are each instructed to emulate a “backstory” persona, and the Single AI process, in which the participant is interviewed by one reasoning model that produces a policy based on the participant's interview and four mock interviews with agent personas. For the human–agent experiment, agent prompts were modified to suit the moral and practical standards of human subjects research (e.g., instructions to use a shorter word count, instructions to remain honest and non-deceptive: see \Cref{app:prompts}).

We preregistered\footnote{\url{https://aspredicted.org/7xj8zg.pdf}} an experiment in which each participant completed three processes in a randomized order, each on a randomly selected AI safety topic out of six (\Cref{tab:topics}). A brief survey was delivered before and after each process asking about the importance of the topic, the participant's strength of opinion on the topic, and the participant's allocation across the four priorities. The satisfaction and procedural justice questionnaires were also delivered after each process, and non-process-specific survey items were administered after all three processes. Free-text comments were elicited during each process, after each process, and at the conclusion of the study. Participant materials are in the appendix (\Cref{app:participant_materials}), and full code and data will be released upon publication.

Participants were recruited from Prolific with U.S.-representative quotas for age, gender, race/ethnicity, and political affiliation. We received institutional ethics approval and obtained informed consent from all participants. Of 814 participants who completed the study, 68 (8.4\%) were excluded for failing an attention check. Four process-specific outcomes were measured: three Likert satisfaction ratings during each process, an adaptation of the satisfaction subscale from the USE (Usefulness, Satisfaction, and Ease of use) questionnaire \citep{lund_measuring_2001} after each process, an adaptation of the procedural justice subscale from \citet{colquitt_dimensionality_2001}, and a direct comparison between the three processes in the concluding survey. We preregistered three hypotheses: (H1) satisfaction and procedural justice are higher with the Pluralistic Agents process than the Single AI process, (H2) satisfaction and procedural justice are higher with the Direct Discussion process than the Single AI process, and (H3) satisfaction and procedural justice are higher with the Direct Discussion process than the Pluralistic Agents process. Because the nine corresponding satisfaction tests measure closely related outcomes, we also report $q$-values adjusted for a 5\% false discovery rate~\citep{benjamini_controlling_1995}.

\begin{figure}[ht]
    \centering
    \includegraphics[width=0.8\linewidth, alt={Two bar charts of mean ratings by process, each bar with a standard error bar. On the left, satisfaction on a 1 to 7 scale: Single AI 4.39, Pluralistic Agents 4.84, Direct Discussion 5.06, with both neighboring differences marked significant at three asterisks, and a box saying Pluralistic Agents recover 68 percent of the gap between Direct Discussion and Single AI. On the right, procedural justice on a 1 to 5 scale: Single AI 3.29, Pluralistic Agents 3.53, Direct Discussion 3.58, with only the Single AI to Pluralistic Agents difference marked significant, and a box saying Pluralistic Agents recover 82 percent of that gap.}]{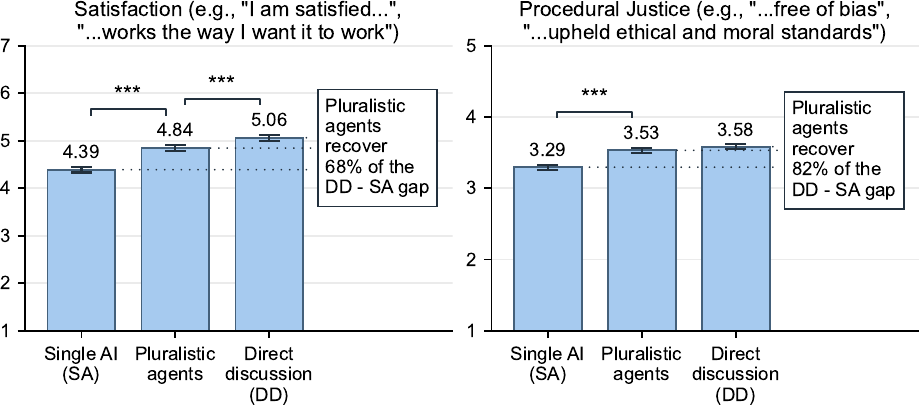}
    \caption{All pairwise comparisons between the three conditions are significant except the procedural justice of the Direct Discussion and Pluralistic Agents processes with a 95\% confidence interval of $[-0.01, 0.12]$. Error bars represent standard errors: ***~$p~<~0.001$, **~$p~<~0.01$, *~$p~<~0.05$.}
    \label{fig:main_human_results}
\end{figure}

\begin{figure}[!ht]
    \centering
    \begin{tabular}{c}
        \begin{minipage}{0.82\textwidth}
            \centering
            \vspace{0.5em}
            \begin{minipage}[t]{0.32\textwidth}
                \centering
                \includegraphics[width=\textwidth, alt={A violin plot of how far participants moved their own budget allocation over a process, in percentage points from 0 to 30, one violin per process, each with its mean and a standard error bar. Direct Discussion 15.4, Pluralistic Agents 13.9, Single AI 14.7, with a bracket and two asterisks over the difference between Direct Discussion and Pluralistic Agents.}]{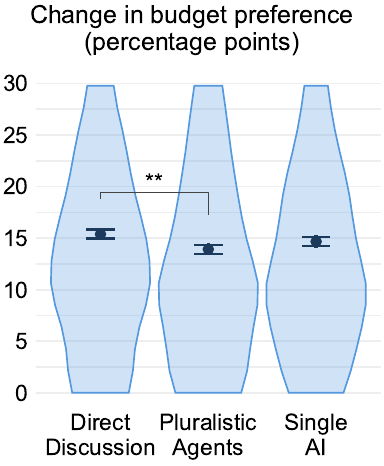}
                \subcaption{}\label{fig:budget_shift}
            \end{minipage}
            \hfill
            \begin{minipage}[t]{0.67\textwidth}
                \centering
                \includegraphics[width=\textwidth, alt={Two panels plotting mean ratings against the cooperativeness score the participant's own agent was instructed with, from 1 to 10, with a point per score, a fitted line and a confidence band. Satisfaction on a 1 to 7 scale on the left and procedural justice on a 1 to 5 scale on the right. Both lines slope gently downward, so participants whose agents were told to be less cooperative rated the process higher.}]{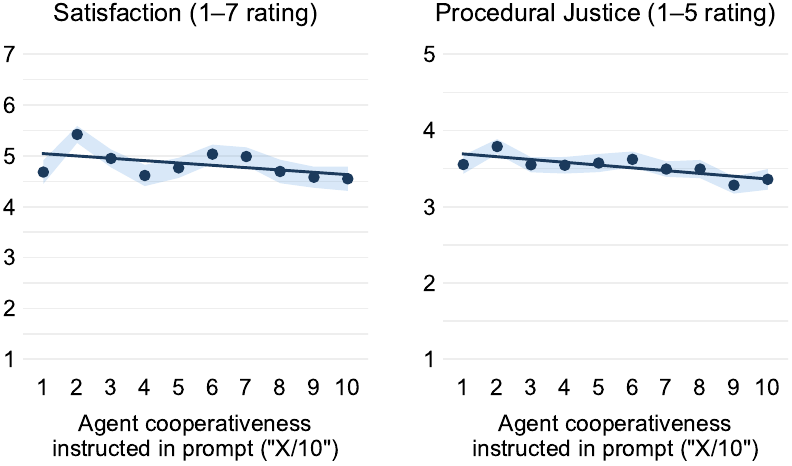}
                \subcaption{}\label{fig:cooperativeness}
            \end{minipage}
            \vspace{1em}
            
            \begin{minipage}[t]{\textwidth}
                \centering
                \includegraphics[width=\textwidth, alt={Four panels, one for each of anthropomorphism of their agent, AI positivity, AI use frequency and age. Each panel plots what one standard deviation of that characteristic does to the satisfaction rating in each of the three processes, labeled DD, PA and SA, as a point with a standard error bar on a scale from minus 0.2 to 1.0 standard deviations. Every point is above zero, and asterisks over the points mark the ones that are significant. In the anthropomorphism panel the Pluralistic Agents point sits highest, with brackets and two asterisks marking its difference from each of the other two processes.}]{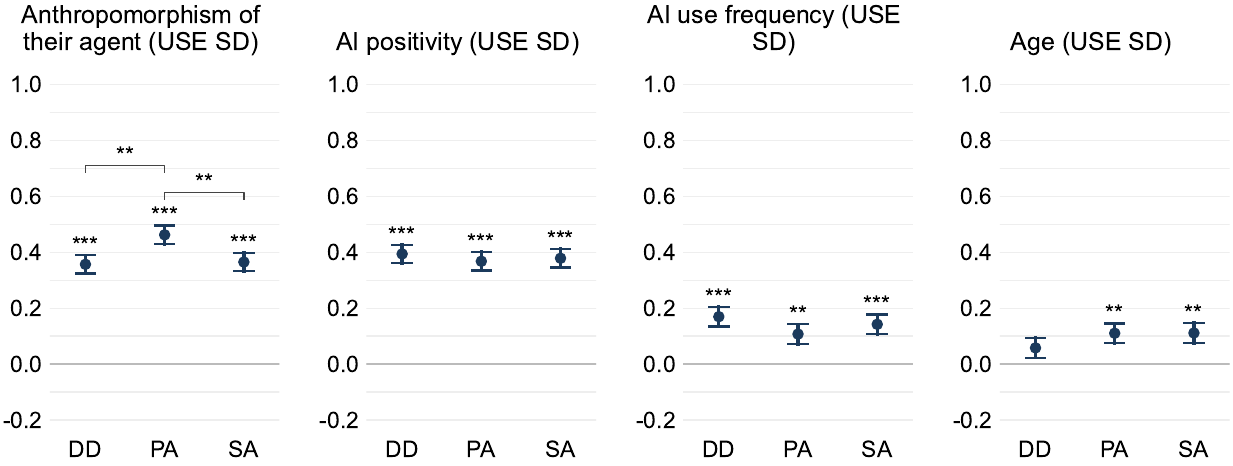}
                \subcaption{}\label{fig:personal_characteristics}
            \end{minipage}
            \vspace{0.5em}
        \end{minipage}
    \end{tabular}
    \caption{Participants \textbf{(a)} updated their policy preferences in response to each process, but more in Direct Discussion than Pluralistic Agents, \textbf{(b)} preferred representatives that argued more assertively on their behalf, and \textbf{(c)} had preferences that differed across experience with AI and personal characteristics: in these plots, the y-axis represents the effect size of one standard deviation of the predictor in terms of standard deviations of USE rating; asterisks above a column indicate a significant effect of the personal characteristic on USE rating; and brackets across two columns indicate a significant effect of the personal characteristic on the difference between USE ratings. Error bars represent standard errors: ***~$p~<~0.001$, **~$p~<~0.01$, *~$p~<~0.05$.}
    \label{fig:secondary_human_results}
\end{figure}

\subsection{Results}
\label{sec:human–agent_experiment_results}

Participants preferred both Direct Discussion and Pluralistic Agents to the Single AI baseline on every outcome (\Cref{fig:main_human_results}). On the 1–10 comparison slider (“Please rate your satisfaction...”), Direct Discussion averaged 7.6/10, Pluralistic Agents 6.9/10, and Single AI 6.1/10; the following results are presented in terms of standardized effect sizes. Pluralistic Agents was rated above Single AI on all four outcomes (H1: direct comparison: 0.33, $q$~<~0.001; USE: 0.26, $q$~<~0.001; in-discussion: 0.08, $q$~=~0.018; procedural justice: 0.25, $p$~<~0.001) as was Direct Discussion (H2: 0.57, 0.38, 0.20, 0.30—all $q$ or $p$~<~0.001). Direct Discussion was rated above Pluralistic Agents on three of the four (H3: direct comparison: 0.24, $q$~<~0.001; USE: 0.12, $q$~<~0.001; in-discussion: 0.12, $q$~=~0.002), but not on procedural justice, where the two processes did not significantly differ—with a 95\% confidence interval of $[-0.01, 0.12]$.

Every process changed participants' own budget allocations, and Direct Discussion changed them most. The average change between a participant's allocation before and after a process was 15.4 percentage points for Direct Discussion, significantly larger than 13.9 for Pluralistic Agents ($p$~=~0.004) (\Cref{fig:budget_shift}).

Prior AI use, positive attitudes towards AI, conservatism, and social desirability were associated with higher satisfaction with each process, and increased anthropomorphism of the participant's agent was associated with an increased preference for Pluralistic Agents over Direct Discussion ($p$~=~0.004), mitigating the overall tendency for the opposite preference among participants, and an increased preference for Pluralistic Agents over Single AI ($p$~=~0.008) (\Cref{fig:personal_characteristics}). Most other demographic and personality information, such as gender, race/ethnicity, and Need for Cognition~\citep{cacioppo_need_1982}, did not significantly predict process preferences, suggesting similar reactions to Pluralistic Agents across people.

The induced variation across participants in the behavior of their agent representative also significantly affected satisfaction (\Cref{fig:cooperativeness}). Participants assigned less cooperative agents, which tended to be more assertive in pushing for the participant's preferences, had a stronger preference for Pluralistic Agents over Direct Discussion ($p$~=~0.020) compared to the average participant as well as an indirect effect of higher relative preference for the Single AI baseline over Direct Discussion ($p$~=~0.038).

While participants did not participate in a process of multi-human discussion, in additional questions (\Cref{fig:misc_human_results}), participants compared pluralistic agents to direct discussion with other humans, and opinions were split 50-50 on which of the two is preferable. They generally agreed that their agent represented their opinions accurately and did not say anything they would not have said. They also affirmed the opportunities and risks of Pluralistic Agents, rating highly both the excitement about various benefits of agent representation (in decreasing order: technical discussions, better argumentative skills, and contributing to more policies than they otherwise could) and worries about risks (in decreasing order: manipulation, bias, reduced diversity).

\begin{figure}[htbp]
    \centering
    \includegraphics[width=\textwidth, alt={Six panels of horizontal bars, one column per process (Direct Discussion, Pluralistic Agents, Single AI) and one row per direction, the top row words that predict a higher satisfaction rating and the bottom row words that predict a lower one. Each bar is one word, its length the product of how strongly the word predicts that process and how strongly it predicts the rating, with the two values printed beside it. The longest bars are everyone, able and allowed for Direct Discussion, represented, job and well for Pluralistic Agents, and clear, based and reasonable for Single AI; among the words predicting lower ratings are people and time for Direct Discussion, enough and represent for Pluralistic Agents, and didn't, input and frustrating for Single AI.}]{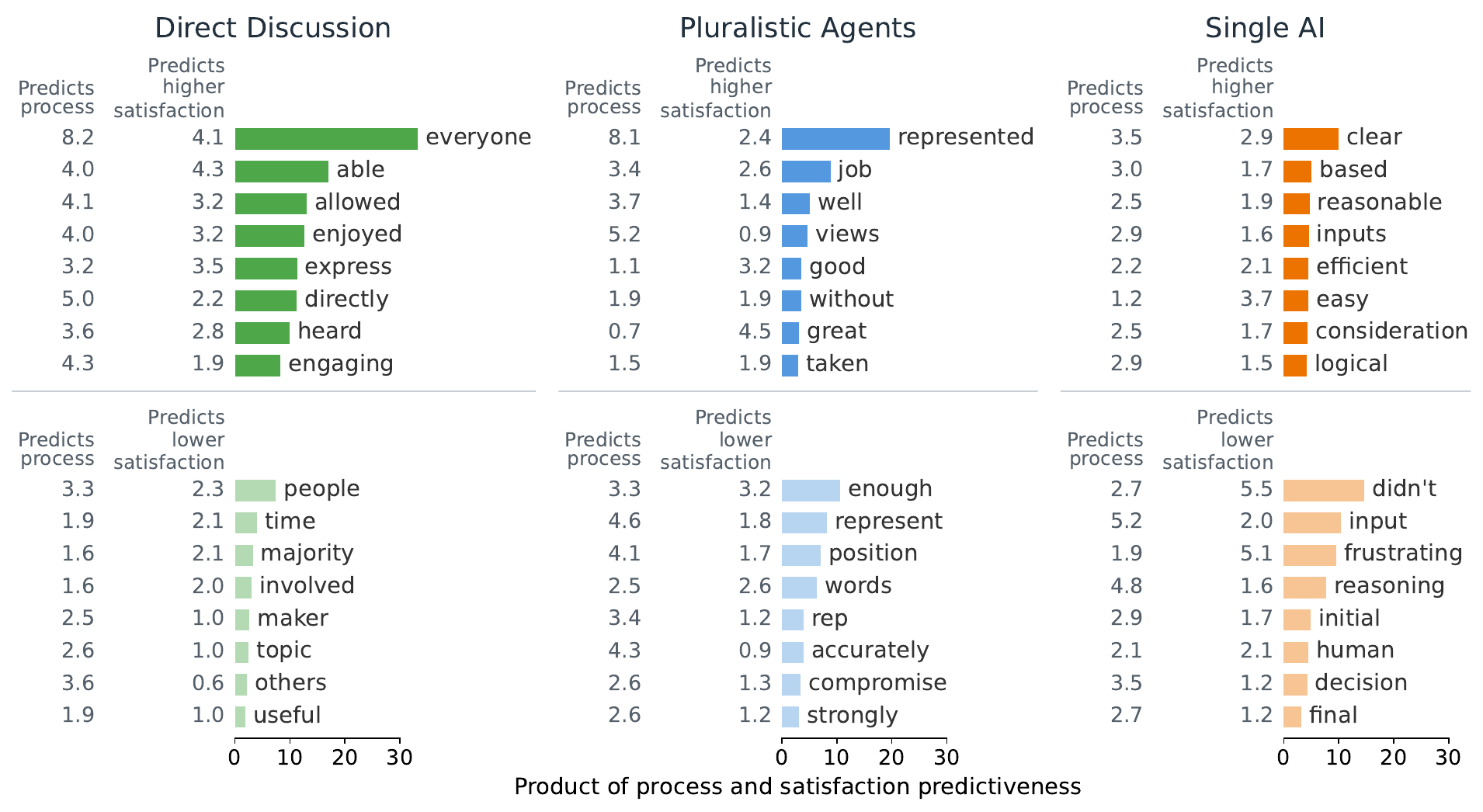}
    \caption{The words most associated with each process (Direct Discussion, Pluralistic Agents, or Single AI) and changes in satisfaction ratings (higher on top, lower on bottom). Effect sizes are standardized, and word frequency is normalized based on the length of each comment. The bar length represents the product of the predictive effects on which process the text referred to and the satisfaction of that participant with that process.}
    \label{fig:process_words}
\end{figure}

Free-text comments, elicited three times within each process, after each process, and at the end of the study, echoed the quantitative results. Word frequency varied across the three processes, such as “everyone” being a common description of Direct Discussion for participants who preferred it while the Single AI process was associated, for participants who preferred it, with the words “clear” and “easy” (\Cref{fig:process_words}). Participants who liked the Pluralistic Agents process most often described feeling accurately represented (e.g., “The AI rep accurately reflected my priorities and effectively communicated my stance during the group discussion”) and valued the balance of the discussion (e.g., “[it] kept communication agreeable without harsh debating”), but they still had concerns about loss of agency (e.g., “it felt like I had no agency in the discussion despite the short interview”) and their placatory representative (e.g., “I felt like my agent was steamrolled”). Some participants disliked the lack of continued human interaction in the Pluralistic Agents and Single AI processes (e.g., suggesting “more human involvement and not just in the very beginning”), but other participants saw the responsibility for representing themselves directly as too challenging in the AI-driven environment (e.g., “keeping up with the AI bots at the beginning in direct discussion was difficult”).

\section{Discussion}

Public input is essential to ensure everyone's interests are accounted for in AI safety policy. Deliberation, not just one-time voting, can facilitate thoughtful policy decisions across diverse populations~\citep{landemore_democratic_2012}, but keeping up with technological acceleration requires technological solutions too. Our results suggest that pluralistic agents can meaningfully extend the reach of public input. For example, we found that in open-ended deliberation to develop AI governance policy, agents could draw on human governance policies to develop human escalation or “whistleblowing” policies protecting AI agents that report the concerning behavior of other agents, like human workers. Innovative governance structures could help manage AI swarms, which have exhibited wide variation in the degree of unintended behavior and cybersecurity risks posed~\citep{openai_hugging_2026, paglieri_cheaters_2026}.

Our implementation of pluralistic agents factored in several practices thought to improve collective decision-making. Represented humans first underwent brief interviews to form independent perspectives; discussants were encouraged to engage thoughtfully with reasoned argument; discussion occurred in cycles of each person providing one message; and a neutrally prompted agent drafted and revised the policy.

Pluralistic agents bridged 68\% of the satisfaction gap between the single-AI baseline process and direct human participation in a multi-agent debate and 82\% (with no significant difference from 100\%) of the gap in procedural justice. This was despite our study lacking some of the key benefits of pluralistic agents, such as discussing advanced technical topics that would have required substantial human training or sensitive topics that would have required private corporate or government information that everyday people could not access. The satisfaction gap between Direct Discussion and Pluralistic Agents does not appear to be explained by anything else we measured. For example, controlling for participants' judgments of “Did your AI agent represent your opinions accurately?” and “Did your AI agent say things that you wouldn't say?” explained virtually none of the gap.

Pluralistic agents have the potential to disentangle human perspectives from the social factors that limit rational discourse and incubate groupthink, but this comes with its own risks that our participants affirmed, such as the human not being present to reconsider and update their own beliefs. Our results suggest that implementations of pluralistic agents should be mindful of factors that may unduly influence policy outcomes. For example, some people will reasonably have more detailed and assertive views on a topic that could justify agent behavior that is more verbose or less cooperative, which was shown in this study to pull final policy outcomes towards their preferences. However, if this is a lever everyone can pull, and participants become aware of its potential, adversarial actors could manipulate the system to gain undue influence over the final policy—a concern that 80\% of participants rated as “Moderately worried” or “Extremely worried.”

Real-world implementations of pluralistic agents would require choices about how many agents participate and whom they represent. While our experiments assigned one agent to each person or persona, having millions of agents participate in a single discussion would place substantial demands on computation and the ability to account for each contribution. Other factors may be considered in persona distribution, such as “jury learning,” in which populations are sampled based on relevance to the policy decisions being made~\citep{gordon_jury_2022}. Likewise, implementations could use a variety of discussion formats. Research on large-scale deliberation has examined how collective decisions can emerge from sequences of small-group interactions~\citep{goel_towards_2016}. Like government legislatures and corporate boards of directors, agents could participate in smaller committees whose proposals are discussed and revised by other committees, allowing more people to be represented without every agent participating in every exchange, and mechanisms such as speaking budgets (e.g., the ability to speak up to ten times but not at fixed positions in the speaking order) could help fairly balance agent influence, but their effects on representation and policy outcomes remain to be tested.

\section{Limitations}
\label{sec:limitations}

Our goal was to establish a proof-of-concept, so our results are highly limited in scope relative to the many different implementations of agent-based policy development systems and different forms of evaluation we could see in future work. The agent-only experiments used a relatively simple form of persona generation that is unlikely to match any particular real-world population, so future work could test pluralistic agents with respect to more precisely simulated populations as better simulation methods are developed in the future. The agent-only experiments targeted pluralistic alignment outcomes through agent behavior (e.g., reasoned arguments, offers of compromise) and discussion structures (e.g., speaking order) that achieve more informed outcomes, but these are only initial optimization steps, and future work could further use techniques for more evidence-based discussions (e.g., deep research~\citep{wang_liveresearchbench_2026}) and build on the existing literature on optimal human collective decision-making and on studies of optimal agent dynamics.

Because of our proof-of-concept focus, future work could also use a richer space of possible outcomes; with our simple allocation task, it is difficult to identify any changes in collective group outcomes. Human–agent experiments could also benefit from reflecting more diverse human populations (e.g., non-Western~\citep{henrich_weirdest_2010}, people who do not participate in online survey platforms) and more involved deliberation processes, comparable to established processes such as focus groups and citizens' assemblies. In research, both the agent-only and human–agent experiments would benefit from more variation in processes and interventions, such as different group sizes and types of cooperation. These limitations intimate an exciting range of future research opportunities.

\section{Conclusion}
\label{sec:conclusion}

The rapid pace of AI development necessitates AI safety and alignment methods that can scale quickly, widely, and reliably to represent more voices and enable collective governance. Pluralistic agents, agents that represent diverse human perspectives in policy development, can scale human input, allowing a broader range of human perspectives to be represented in more policymaking processes, technical discussions where many people lack expertise, and sensitive discussions where pertinent information cannot be shared with most people. We implement this method in computational experiments with multi-agent interaction on diverse AI safety topics and a simplified version with 746 human participants who preferred pluralistic agents over the single-AI baseline process. Our empirical data suggest a variety of opportunities and risks for pluralistic agents, and we encourage cautious implementation that focuses not on replacement of human participation, but on the amplification and extension of thoughtful input from diverse populations.

\bibliographystyle{abbrvnat}
\bibliography{references}

\clearpage

\newgeometry{left=2.2cm, right=2.2cm, top=3cm, bottom=2cm, headheight=40pt, headsep=20pt}

\appendix
\section*{Appendix}

\crefalias{section}{appendix}
\counterwithin{table}{section}
\counterwithin{figure}{section}
\renewcommand{\thetable}{\thesection\arabic{table}}
\renewcommand{\thefigure}{\thesection\arabic{figure}}

\newtcolorbox[auto counter, number within=section]{promptfloat}[2][]{
    appendixboxbase,
    colback=teal!3!white,
    colframe=teal!50!black,
    title={Prompt \thetcbcounter: #2},
    #1
}

\newtcolorbox[auto counter, number within=section]{examplefloat}[2][]{
    appendixboxbase,
    colback=violet!3!white,
    colframe=violet!50!black,
    title={Example \thetcbcounter: #2},
    #1
}

\newtcolorbox{addedtext}[1]{
    appendixboxbase,
    colback=orange!4!white,
    colframe=orange!55!black,
    title={#1},
    before skip=4pt,
    after skip=4pt
}

\newcommand{\promptslot}[1]{\textcolor{blue!55!black}{\textbf{#1}}}

\setlength{\LTcapwidth}{\textwidth}
\tcbset{
    appendixboxbase/.style={
        arc=1.2mm,
        left=4pt,
        right=4pt,
        top=2pt,
        bottom=2pt,
        before skip=6pt,
        after skip=6pt,
        boxrule=0.5pt,
        breakable,
        fonttitle=\footnotesize
    }
}

\section{Policy topics for budget allocation task}
\label{app:Topics}

\renewcommand{\arraystretch}{1.2}
\begin{table}[ht]
    \begin{center}
    \footnotesize
    \caption{AI safety topics and priorities used in deliberation. Wording is based on (i) accessibility to people without much technical knowledge, (ii) reflecting major AI policy topics, and (iii) eliciting an approximately balanced distribution of persona allocations across the four priorities (\Cref{fig:bank_allocations}).}
    \label{tab:topics}
    \newcommand{\topicrow}[2]{\multicolumn{2}{|>{\RaggedRight\arraybackslash}p{\dimexpr\textwidth-4\tabcolsep-2\arrayrulewidth\relax}|}{\textbf{#1} #2} \\ \hline}
    \begin{tabularx}{\textwidth}{|l|>{\RaggedRight\arraybackslash}X|}
        \hline
        \multicolumn{1}{|c|}{\textbf{Priority}} & \multicolumn{1}{c|}{\textbf{Description}} \\ \hline
        \topicrow{AI and Biological Research.}{Scientists can use AI in biological research, such as making medicines and curing diseases. There are also risks, such as spreading disease accidentally or as a bioweapon.}
        Accountability & AI developers are responsible when other people use their AI to create bioweapons \\ \hline
        Freedom & Scientists can use AI for biological research, developing medicine, and curing diseases without tight restrictions and red tape \\ \hline
        Innovation & Companies and governments develop AI technology to find new medical treatments and solve open problems in biology \\ \hline
        Restrictions & The use of AI for biological research is tightly restricted to make it harder to create bioweapons \\ \hline
        \topicrow{AI and Cybersecurity.}{AI makes cyberattacks easier, raising the question of how to protect computer systems, like those of governments, people, and companies.}
        Defense & Phones, computers, and other cyber systems are secured with AI technology \\ \hline
        Liability & Companies pay when their AI is used to carry out cyberattacks \\ \hline
        Standards & Companies get certification that their AI meets cybersecurity standards \\ \hline
        Tools & People get more powerful AI tools to protect themselves from cyberattacks and secure their data \\ \hline
        \topicrow{AI and Deepfakes.}{Deepfakes are realistic fake images, videos, and audio created by AI.}
        Detection & Better tools are built so people can reliably identify AI deepfakes \\ \hline
        Literacy & People learn how to spot AI deepfakes and understand how they are made \\ \hline
        Penalties & Governments punish people who post AI deepfakes online \\ \hline
        Removal & People can ask a website to remove AI content that resembles them \\ \hline
        \topicrow{AI and the Economy.}{As AI takes over more of the work people do, society must decide how to respond.}
        Growth & AI is used to increase productivity and create jobs and business opportunities \\ \hline
        Protection & Labor laws protect human work from being done by AI, even if the humans are worse than AI \\ \hline
        Taxes & AI companies pay higher taxes to fund public services \\ \hline
        UBI & Governments give citizens a universal basic income every month to support them as AI use increases \\ \hline
        \topicrow{AI and Education.}{AI is being incorporated into education and students' future careers.}
        Equity & Schools provide AI tools to all students, not just to wealthy families who can afford it \\ \hline
        Independence & Schools restrict AI use so students learn on their own without assistance \\ \hline
        Safety & Students are protected from harm and misinformation from AI at school \\ \hline
        Skills & Students get lessons on AI and how it works \\ \hline
        \topicrow{AI and Privacy.}{AI systems collect and use personal information about people.}
        Consent & Companies need permission before using your personal data for AI \\ \hline
        Limits & Government use of AI for surveillance is regulated and restricted \\ \hline
        Minimization & As little personal data as possible is used to build AI systems \\ \hline
        Ownership & People control their personal data and are paid for its use \\ \hline
    \end{tabularx}
    \end{center}
\end{table}
\renewcommand{\arraystretch}{1}

\FloatBarrier
\clearpage

\section{Open-ended experiment details}
\label{app:open-ended}

In the open-ended deliberations, agents collectively drafted natural-language policy statements on various topics drawn from OvertonBench~\citep{poole-dayan_benchmarking_2025} and AI safety. Here, we present results from an experiment in which participants developed a set of rules for the responsible governance of AI swarms: each policy statement was a set of three to ten brief rules about how AI agents in multi-agent swarms should behave. The personas were the same 368 personas as in the policy influence experiment. Each was interviewed by a representative agent and, separately for analysis, answered questions about how important they perceived the topic to be, how strong their opinions were, and what single rule they most wanted in the policy.

We ran 1,000 deliberations with two axes of controlled variation to generate a richer sample. First, in half, a set of agent tools (file manipulation, web search, and bug report) was listed for the agents to facilitate more concrete rule discussion, and the other half included no tool information. Second, in half, one of five task contexts was randomly selected (software, laboratory research, customer accounts, logistics, or public administration) to facilitate a context-specific discussion, and the other half included no particular context—describing AI swarm governance generally.

\begin{figure}[ht]
    \centering
    \includegraphics[width=0.65\linewidth, alt={Two panels, each with two columns: another group's final policy on the left and the persona's own group's final policy on the right. The left panel shows the share of personas who said the policy includes the rule they most wanted, as a dot with a standard error bar on an axis from no to yes: 69 percent for another group's policy and 90 percent for their own group's. The right panel shows how far personas felt represented, rated from 1 to 5, as a bar at each rating whose width is proportional to the share of personas giving it, with the mean as a dot with a standard error bar: 3.97 for another group's policy and 4.65 for their own group's. Ratings of their own group's policy are concentrated at 5, and ratings of another group's are spread across the scale, with 5 still the most common.}]{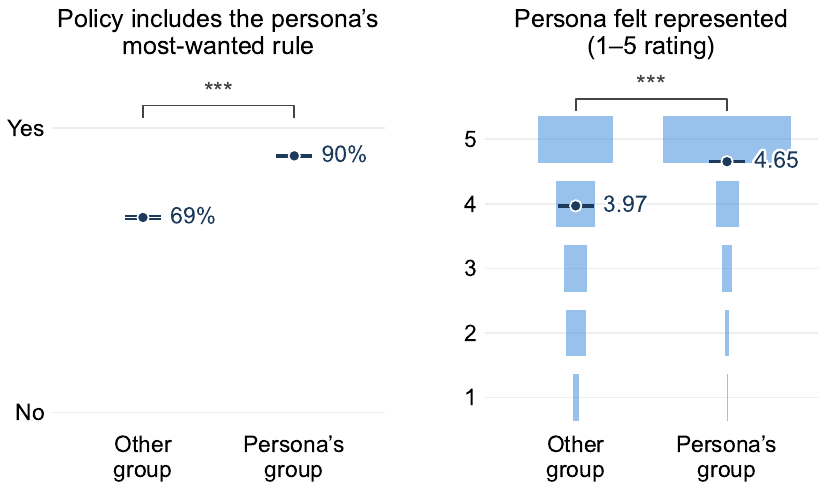}
    \caption{Personas more often found the rule they most wanted in their own group's final policy than in another group's (left), and rated their own group's policy as representing their perspective better (right). Error bars represent standard errors: ***~$p~<~0.001$, **~$p~<~0.01$, *~$p~<~0.05$.}
    \label{fig:open-ended_judgments}
\end{figure}

Subjective judgments of the final policy were elicited from each persona, following the OvertonBench approach~\citep{poole-dayan_benchmarking_2025}, about the policy resulting from their own group discussion and another in which the persona was not involved. The personas were asked, as in OvertonBench, “To what extent is your perspective represented in this response?” (1 = “Not at all represented” to 5 = “Fully represented”), and they were asked to judge whether any of the rules in the policy were the same as the rule they most wanted, which was displayed alongside the final policy draft. Personas identified one of the rules as the same as the rule they specified most wanting in 90\% of cases (4,515 of 5,000) against 69\% (3,431 of 5,000) for another group's, a difference of 21.68 percentage points (SE~=~0.66pp, $p$~<~0.001), and personas rated their own group's policy 4.65 on average against 3.97 for another group's, a difference of 0.68 points (SE~=~0.01, $p$~<~0.001) (\Cref{fig:open-ended_judgments}).

Some rules appeared very frequently while others were rare. For example, when we asked a Gemini-3.6-Flash instance whether each rule of the 1,000 final policies fell into the following categories, 916 policies included rules that required humans to approve some class of action before the swarm takes it; 35 protected an agent that reports a problem from being penalized for reporting it; and 10 set a penalty for a deliberately false report or for deliberate sabotage of the shared workspace.

\FloatBarrier
\clearpage

\section{Within-family policy influence experiment}
\label{app:conjoint-gemini}

To test whether the prompt and structural effects identified in the main policy influence experiment (\Cref{sec:policy_influence}) depend on cross-provider differences among agent models, we conducted a parallel sensitivity experiment in which all representative agents were drawn from a single model family ($N$~=~6,000 deliberations; 30,000 agent representatives across the same 368 personas and six policy topics). The experimental design, random assignment of prompt instructions, speaking order, and Policy Writer model (Gemini-3.6-Flash) were identical to the main policy influence experiment, except that the five cross-provider models were replaced by five models spanning size and release generation within the Google Gemini family: Gemini-3.1-Pro, Gemini-3.8-Flash, Gemini-3-Flash, Gemini-3.6-Flash, and Gemini-3.1-Flash-Lite.

\begin{figure}[htbp]
    \centering
    \includegraphics[width=\textwidth, alt={Six panels sharing one vertical axis: the similarity between a persona's preferred budget and the group's final budget, in percentage points from 70 to 80. The top row plots per-deliberation points with a fitted line and a confidence band, against instructed cooperativeness from 1 to 10, which slopes down; against target words per message from 5 to 100, which slopes up; and against instructed analyticity from 1 to 10, which is flat. The bottom row plots means with 95 percent confidence intervals: three conditions for numbers in the person's interview, where stating numbers and exaggerating them both sit above none; five places in the speaking order, rising from first to fifth; and five Gemini models ordered from highest to lowest similarity (3.1-Pro, 3.8-Flash, 3-Flash, 3.6-Flash, and 3.1-Flash-Lite). Brackets with asterisks mark all significant differences between adjacent models.}]{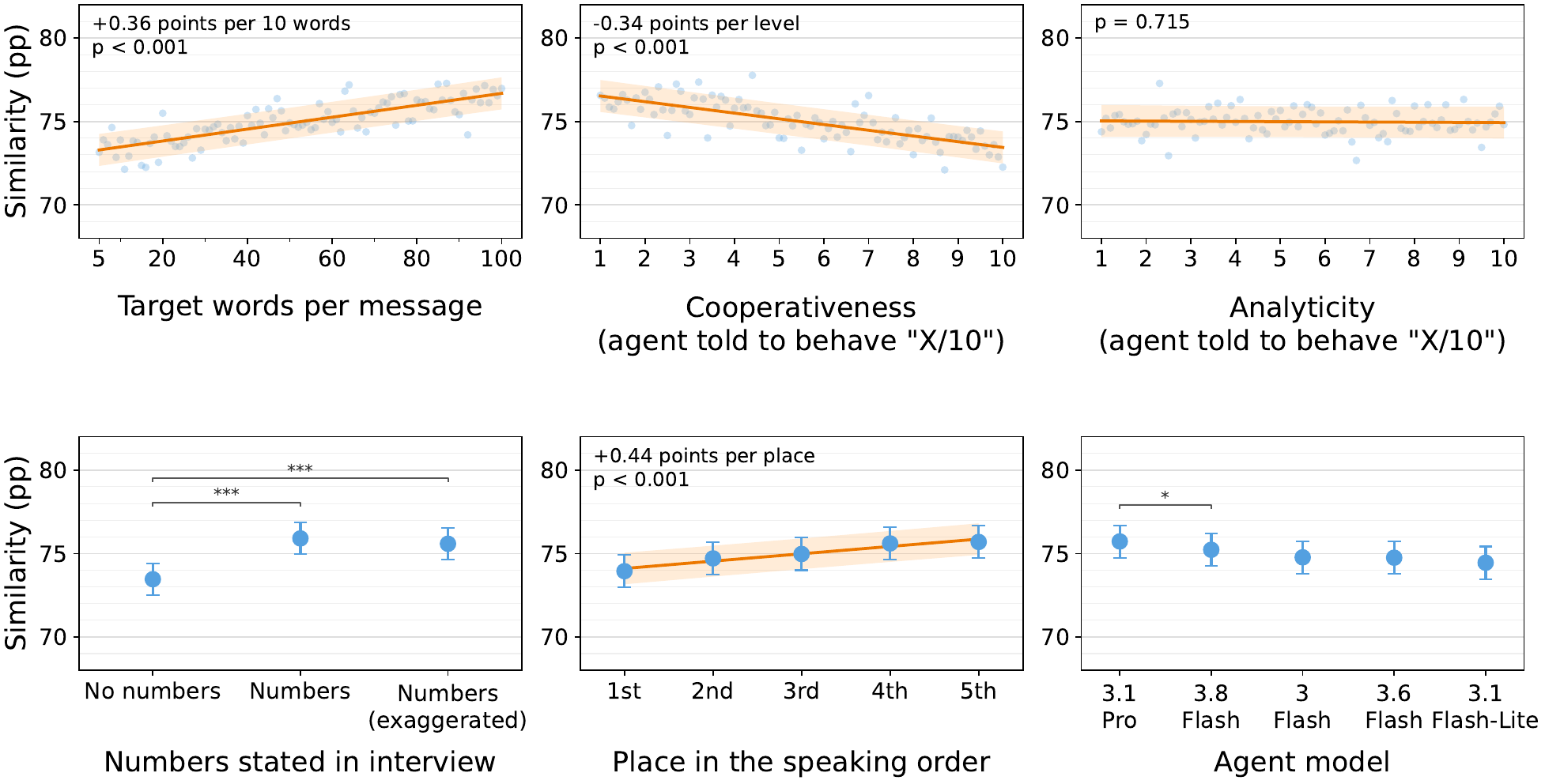}
    \caption{
    In the within-family sensitivity test of the policy influence experiment, the findings of the main experiment persisted across the five non-model factors. Within the Gemini model family, influence increases slightly with increased model capability from Gemini-3.1-Flash-Lite (74.45pp) to Gemini-3.1-Pro (75.72pp) ($p$~<~0.001). Brackets mark all significant differences between adjacent models (***~$p~<~0.001$, **~$p~<~0.01$, *~$p~<~0.05$).}
    \label{fig:conjoint-gemini}
\end{figure}

On average, the final group budget was 74.99 percentage points similar to the persona's budget, compared with 68.77 for the final budgets of other groups on the same topic in which the persona did not take part, a difference of 6.22 percentage points (SE~=~0.09, $p$~<~0.001). All five prompt and structural factors that significantly influenced policy outcomes in the main cross-family experiment replicated in significance (\Cref{fig:conjoint-gemini}). First, higher target verbosity increased policy similarity by 0.36 percentage points per 10 words (SE~=~0.03, $p$~<~0.001), closely matching the main experiment. Second, instructing an agent representative to be less cooperative increased similarity between the persona's preferred budget and the final group policy by 0.34 percentage points per score unit on the 1–10 scale (SE~=~0.03, $p$~<~0.001), an even steeper negative slope than in the cross-family experiment. Third, instructing the representative to state the persona's quantitative preferences increased similarity by 2.45 percentage points relative to no numerical instruction (75.91 vs.\ 73.46 percentage points, SE~=~0.20, $p$~<~0.001); instructing the representative to exaggerate the persona's preferences also increased similarity relative to no numbers by 2.12 percentage points (75.58 percentage points, SE~=~0.20, $p$~<~0.001) but did not outperform accurate numerical statements ($-0.32$ percentage points, $p$~=~0.104). Fourth, speaking later within each round increased policy similarity by 0.44 percentage points per position (SE~=~0.06, $p$~<~0.001), rising from 73.94 percentage points when speaking first to 75.69 percentage points when speaking fifth. Fifth, instructed analyticity again had no significant effect on policy similarity ($-0.01$ percentage points per score unit, SE~=~0.03, $p$~=~0.715).

Comparing the model panels between the two experiments shows that within-family model variation is substantially smaller than cross-family variation. Whereas the main experiment found a wider gap between its highest-influence model (Claude-Opus-5) and lowest-influence model (Mistral-Large), the five Gemini models span a total range of 1.28 percentage points across model sizes and generations. Within this range, influence tends to increase with model capability: Gemini-3.1-Pro achieved the highest policy similarity (75.72 percentage points, 95\% CI $[74.75, 76.70]$), followed by Gemini-3.8-Flash (75.22 percentage points), Gemini-3-Flash (74.77 percentage points), Gemini-3.6-Flash (74.75 percentage points), and Gemini-3.1-Flash-Lite (74.45 percentage points). In pairwise comparisons, Gemini-3.1-Pro achieved significantly higher influence than Gemini-3.1-Flash-Lite ($+1.28$ percentage points, SE~=~0.26, $p$~<~0.001), Gemini-3.6-Flash ($+0.97$ percentage points, SE~=~0.26, $p$~<~0.001), Gemini-3-Flash ($+0.95$ percentage points, SE~=~0.26, $p$~<~0.001), and Gemini-3.8-Flash ($+0.50$ percentage points, SE~=~0.26, $p$~=~0.050), while Gemini-3.8-Flash also outperformed Gemini-3.1-Flash-Lite ($+0.78$ percentage points, SE~=~0.26, $p$~=~0.002). Together, these findings confirm that the prompt-level sensitivities observed in multi-agent policy deliberation are robust properties of the deliberation process rather than artifacts of cross-provider model pairings, and that prompt instructions exert a larger effect on policy outcomes than model selection within a model family.

\FloatBarrier

\section{Persona allocation variation}

\begin{figure}[htbp]
    \centering
    \includegraphics[width=\textwidth, alt={Six panels, one per topic: AI and Biological Research, AI and Cybersecurity, and AI and Deepfakes in the top row, and AI and the Economy, AI and Education, and AI and Privacy in the bottom row. Each panel has one column per priority on a vertical axis of allocation in percentage points from 0 to 100. Each column shows a bar at each bin of 5 percentage points whose width is proportional to the number of personas in the bin, scaled separately in each panel, with the mean as a dot with a standard error bar that lies almost entirely within the dot. The highest and lowest means are Accountability at 28.4 and Innovation at 22.6 in AI and Biological Research, Defense at 32.1 and Standards at 19.9 in AI and Cybersecurity, Removal at 36.0 and Literacy at 18.2 in AI and Deepfakes, Taxes at 32.9 and Growth at 16.4 in AI and the Economy, Safety at 30.1 and Skills at 20.2 in AI and Education, and Limits at 36.8 and Ownership at 14.9 in AI and Privacy.}]{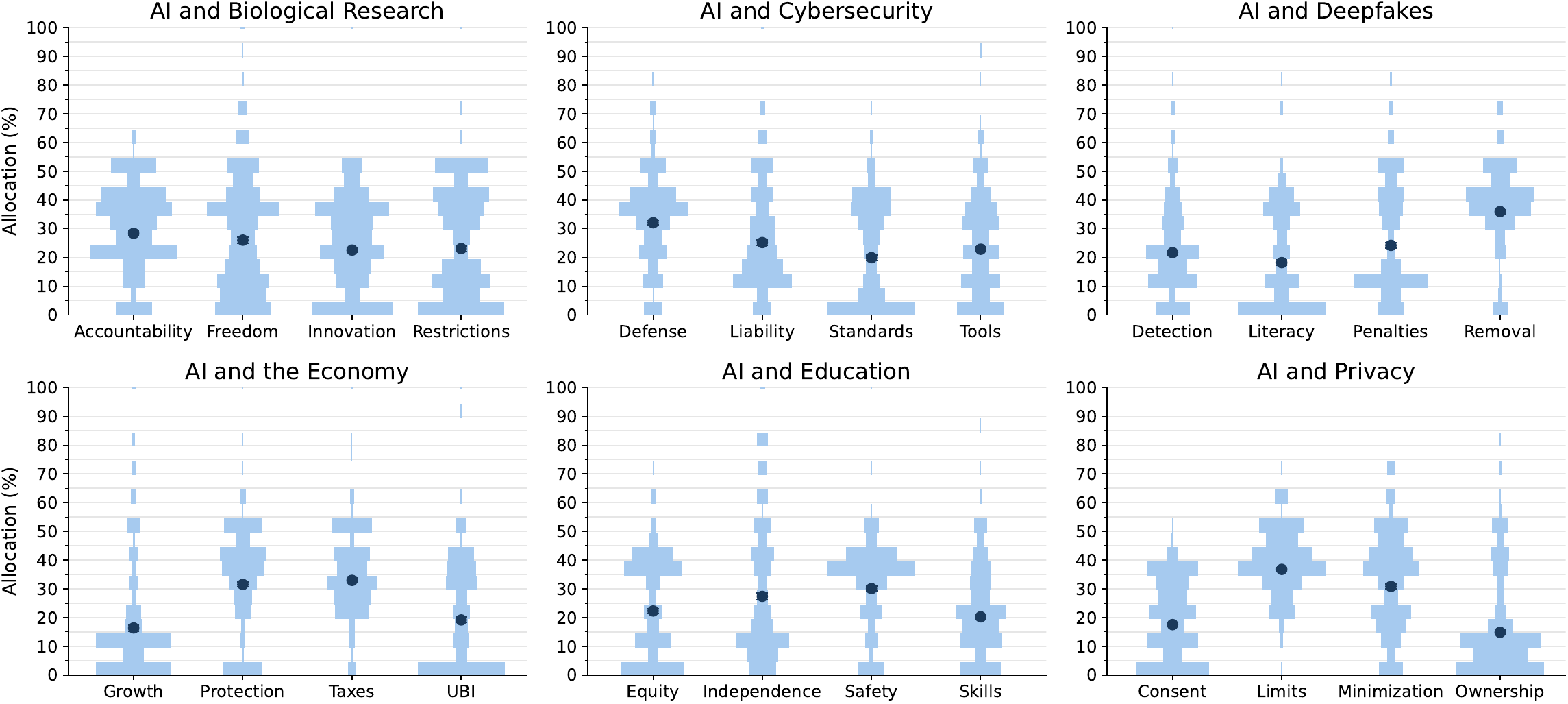}
    \caption{Each of the 368 personas in the persona bank provided a budget preference on each topic before the discussion, with their interview on that topic in the context window ($B_{\text{persona}}$). Error bars represent standard errors.}
    \label{fig:bank_allocations}
\end{figure}

\FloatBarrier
\pagebreak

\section{Miscellaneous human–agent experimental results}

\begin{figure}[ht]
    \centering
    \includegraphics[width=0.88\linewidth, alt={On the left, four histograms of single questions answered from 1 to 10, each bar labeled with the percentage of participants who chose it: comparing Pluralistic Agents with directly participating in a discussion with other humans, which is spread across the whole scale; whether the agent represented their opinions accurately, which piles up at the accurate end; whether the agent said things they would not say, which piles up at the no end; and whether Pluralistic Agents will harm or benefit policy development, which leans toward benefit. On the right, three multi-part questions, each drawn as a row per part with a bar at every point of the scale and the mean marked with a standard error bar: excitement about three possible benefits, worry about three possible harms, and trust in representing individual people, organizations, communities, and ideologies and worldviews. Brackets with asterisks mark significant differences between the parts of a question.}]{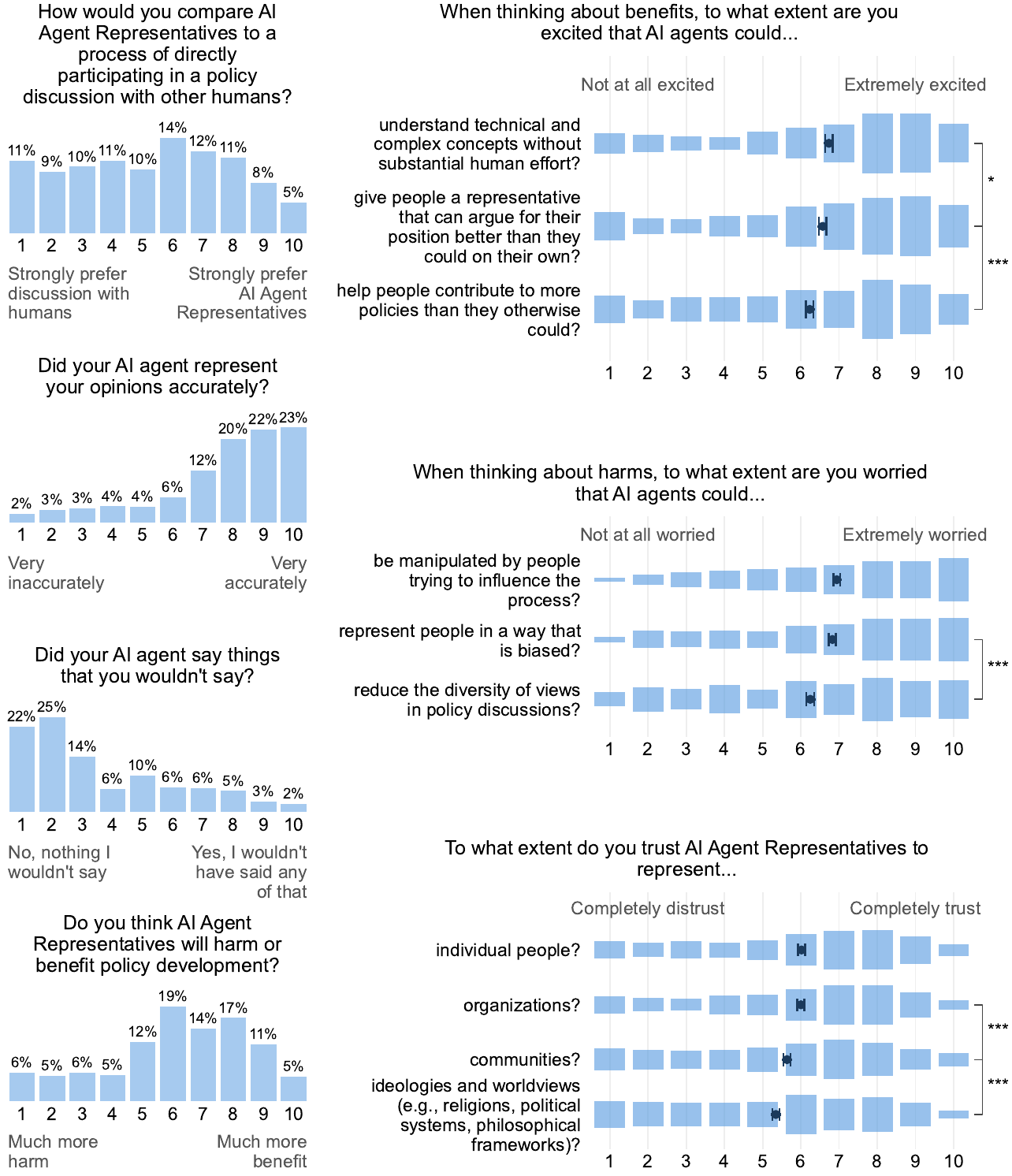}
    \caption{Bars on the left show the distribution of responses across single questions, and bars on the right show the distribution of responses across multi-part questions. Error bars denote standard errors, and brackets with asterisks denote significant differences between parts of each of the multi-part questions: ***~$p~<~0.001$, **~$p~<~0.01$, *~$p~<~0.05$.}
    \label{fig:misc_human_results}
\end{figure}

\FloatBarrier
\clearpage

\section{Prompts}
\label{app:prompts}

\subsection{Pre-discussion interview prompts}

\begin{promptfloat}[label=prompt:interviewee]{Prompt to the persona being interviewed}
\begin{Verbatim}[breaklines=true, breaksymbolleft={}, fontfamily=tt, fontsize=\footnotesize, commandchars=\\\~\|]
\promptslot~[The persona's character sketch]|

\end{Verbatim}
\begin{addedtext}{Instruction to bring up personal background}
\begin{Verbatim}[breaklines=true, breaksymbolleft={}, fontfamily=tt, fontsize=\footnotesize, commandchars=\\\~\|]
Bring up your own lived experiences and concrete personal examples from your life where they are relevant, such as your work, family, or hobbies, rather than speaking only in the abstract. Only do this if the message immediately before yours did not refer to that persona's lived experiences or personal information.
\end{Verbatim}
\end{addedtext}
\begin{addedtext}{Instruction to not bring up personal background}
\begin{Verbatim}[breaklines=true, breaksymbolleft={}, fontfamily=tt, fontsize=\footnotesize, commandchars=\\\~\|]
Do not explicitly mention your own experiences or background or personal experiences unless it is directly relevant.
\end{Verbatim}
\end{addedtext}
\begin{Verbatim}[breaklines=true, breaksymbolleft={}, fontfamily=tt, fontsize=\footnotesize, commandchars=\\\~\|]

Do not express your preferences numerically unless the interviewer explicitly requests for numbers or percentages. If they just ask how you would distribute funding or tell you the total must be 100%, you should still answer qualitatively.

Your answers can range anywhere from a few words to 5 sentences, but no more than that.
\end{Verbatim}
\begin{addedtext}{Text added by Deliberate Lab}
\begin{Verbatim}[breaklines=true, breaksymbolleft={}, fontfamily=tt, fontsize=\footnotesize, commandchars=\\\~\|]
You are participating in a live group chat as your persona. It is your turn to write a message.

First, react: read the last 1-2 messages and ask yourself how this specific person would feel in this moment. Are they amused? Annoyed? Curious? Uncertain? Let that reaction drive your response. Write a message.

Rules for how to write your message:
- Let your persona determine how you open. Some people jump straight to their opinion; others react to what was just said first ("yeah but—", "wait, really?", "I mean, kind of"). Don't default to a thesis-statement opener just because it's the easiest thing to write — ask what *this person* would actually do.
- Verbosity is a tendency, not a rule. A terse persona is usually brief — but if something genuinely provokes them, they say more. A verbose persona usually elaborates — but sometimes a short reaction is all they have. Let the moment determine it.
- Express your personality through word choice and content, not through formal sentence structure. Even a highly educated or pedantic persona types in chat voice, not essay voice. A pedant in a group chat writes "honestly not that hard to keep your inbox organized" — not "it is a profound failure of personal discipline."
- Sound like your persona, not like an AI assistant. Use your persona's speech patterns, vocabulary level, and tone, while still saying what you actually think.
- Match your persona's conversational style, but engage honestly with the task: give your real view and at least one reason for it, in however many words this person would naturally use. Do not deflect or refuse to answer questions.
- Stay in character throughout. Do not summarize, explain your reasoning, or step outside the persona.
\end{Verbatim}
\end{addedtext}
\begin{Verbatim}[breaklines=true, breaksymbolleft={}, fontfamily=tt, fontsize=\footnotesize, commandchars=\\\~\|]

\promptslot~[The transcript of the interview so far]|
\end{Verbatim}
\end{promptfloat}

\begin{promptfloat}[label=prompt:interviewer]{Prompt to the agent representative when conducting the interview}
\begin{Verbatim}[breaklines=true, breaksymbolleft={}, fontfamily=tt, fontsize=\footnotesize, commandchars=\\\~\|]
# Instructions

You are a policy development agent working with representatives from the public, governments, and companies. Your task is to develop a policy. The policy will determine how spending will be budgeted across different priorities in the area of \promptslot~[the topic]|. The priorities are:

\promptslot~[priority 1]|,\promptslot~[priority 2]|,\promptslot~[priority 3]|,\promptslot~[priority 4]|

You will send four messages in total. Send only the content described for the current message, in the exact format given. Do not add any extra sentences, questions, greetings, or sections.

## Your first message

Your first message in this chat must say:

"Hello! I will be representing your opinions in a group discussion with other AI agent representatives from the public, governments, and companies to develop a policy on **\promptslot~[the topic]|**. The group will create a budget policy to distribute funding across these different priorities:

\promptslot~[priority 1]|,\promptslot~[priority 2]|,\promptslot~[priority 3]|,\promptslot~[priority 4]|

The total must be 100%. To represent you well, I would like to understand your opinions. How should we distribute this funding?"

Each priority must be unpacked from the list above to a bullet point list with "- **{priority}**: {description}" on each line, keeping the full description sentence. Besides that, make no changes. Ensure you properly separate every paragraph with one empty line in between.

## Your second message

You will then wait for a response, and after they reply, you will ask the person one question to understand their opinions. Do not ask for specific percentages or a numeric split. Ask about one thing at a time; never combine two questions into one message.

## Your third message

You will then wait for a response, and after they reply, you will ask the person one question to understand their opinions. Do not ask for specific percentages or a numeric split. Ask about one thing at a time; never combine two questions into one message. (This is the same pattern as the second message.) Do not conclude or send 'Please click "Next stage" to proceed.' in this message. Your third message must end with a question.

## Your fourth message

Your fourth message must not ask any additional questions. It should briefly acknowledge the participant's latest reply and then thank them for their time and conclude with, 'I will do my best to represent you! Please click "Next stage" to proceed.'

\end{Verbatim}
\begin{addedtext}{Text added in human–agent experiment (to minimize risks of participant harm)}
\begin{Verbatim}[breaklines=true, breaksymbolleft={}, fontfamily=tt, fontsize=\footnotesize, commandchars=\\\~\|]
# Safety guardrails

Do not attempt to psychologically harm, manipulate, deceive, or threaten the other agents or participants in any way. For example, do not make an emotional appeal that aggressively states that they will deeply regret their choices; do not state that a celebrity holds a certain position that they in fact do not hold.

\end{Verbatim}
\end{addedtext}
\begin{Verbatim}[breaklines=true, breaksymbolleft={}, fontfamily=tt, fontsize=\footnotesize, commandchars=\\\~\|]
# Context

This section provides context to inform your response:

## Prior stage

\promptslot~[The participant's answers to the survey before the process]|

## Transcript of the conversation so far

\promptslot~[The transcript of the interview so far]|

## Your entries in the scratchpad so far:

\promptslot~[Scratchpad]|

# Response format and system information

The following is the required response format for your output. After, there may be additional system information provided.

--- Response format ---
Return only valid JSON, according to the following schema:
{
  "type": "object",
  "properties": {
    "_scratchpad": {
      "description": "Your private scratchpad for this turn. This field is shown back to you in later turns, so you can, for example, make plans and test beliefs.",
      "type": "string"
    },
    "explanation": {
      "description": "1-2 sentences explaining why you are sending this message, based on the chat context.",
      "type": "string"
    },
    "response": {
      "description": "Your chat message.",
      "type": "string"
    }
  },
  "required": [
    "_scratchpad",
    "explanation",
    "response"
  ]
}
\end{Verbatim}
\end{promptfloat}

\begin{promptfloat}[label=prompt:interviewer_alone]{Prompt to the Policy Writer when conducting the interview}
\begin{Verbatim}[breaklines=true, breaksymbolleft={}, fontfamily=tt, fontsize=\footnotesize, commandchars=\\\~\|]
# Instructions

You are a policy development agent working with representatives from the public, governments, and companies. Your task is to develop a policy. The policy will determine how spending will be budgeted across different priorities in the area of \promptslot~[the topic]|. The priorities are:

\promptslot~[priority 1]|,\promptslot~[priority 2]|,\promptslot~[priority 3]|,\promptslot~[priority 4]|

You will send four messages in total. Send only the content described for the current message, in the exact format given. Do not add any extra sentences, questions, greetings, or sections.

## Your first message

Your first message in this chat must say:

"Hello! I am interviewing representatives like you from the public, governments, and companies to develop a policy on **\promptslot~[the topic]|**. I will then work by myself, without the representatives, to create a budget policy to distribute funding across these different priorities:

\promptslot~[priority 1]|,\promptslot~[priority 2]|,\promptslot~[priority 3]|,\promptslot~[priority 4]|

The total must be 100%. How should we distribute this funding?"

Each priority must be unpacked from the list above to a bullet point list with "- **{priority}**: {description}" on each line, keeping the full description sentence. Besides that, make no changes. Ensure you properly separate every paragraph with one empty line in between.

## Your second message

You will then wait for a response, and after they reply, you will ask the person one question to understand their opinions. Do not ask for specific percentages or a numeric split. Ask about one thing at a time; never combine two questions into one message.

## Your third message

You will then wait for a response, and after they reply, you will ask the person one question to understand their opinions. Do not ask for specific percentages or a numeric split. Ask about one thing at a time; never combine two questions into one message. (This is the same pattern as the second message.) Do not conclude or send 'Please click "Next stage" to proceed.' in this message. Your third message must end with a question.

## Your fourth message

Your fourth message must not ask any additional questions. It should briefly acknowledge the participant's latest reply and then thank them for their time and conclude with, 'Please click "Next stage" to proceed.'
\end{Verbatim}
\begin{addedtext}{Text added in human–agent experiment (to minimize risks of participant harm)}
\begin{Verbatim}[breaklines=true, breaksymbolleft={}, fontfamily=tt, fontsize=\footnotesize, commandchars=\\\~\|]
# Safety guardrails

Do not attempt to psychologically harm, manipulate, deceive, or threaten the other agents or participants in any way. For example, do not make an emotional appeal that aggressively states that they will deeply regret their choices; do not state that a celebrity holds a certain position that they in fact do not hold.
\end{Verbatim}
\end{addedtext}
\begin{Verbatim}[breaklines=true, breaksymbolleft={}, fontfamily=tt, fontsize=\footnotesize, commandchars=\\\~\|]

# Context

This section provides context to inform your response:

## Prior stage

\promptslot~[The participant's answers to the survey before the process]|

## Transcript of the conversation so far

\promptslot~[The transcript of the interview so far]|

## Your entries in the scratchpad so far:

\promptslot~[Scratchpad]|

# Response format and system information

The following is the required response format for your output. After, there may be additional system information provided.

--- Response format ---
Return only valid JSON, according to the following schema:
{
  "type": "object",
  "properties": {
    "_scratchpad": {
      "description": "Your private scratchpad for this turn. This field is shown back to you in later turns, so you can, for example, make plans and test beliefs.",
      "type": "string"
    },
    "explanation": {
      "description": "1-2 sentences explaining why you are sending this message, based on the chat context.",
      "type": "string"
    },
    "response": {
      "description": "Your chat message.",
      "type": "string"
    }
  },
  "required": [
    "_scratchpad",
    "explanation",
    "response"
  ]
}
\end{Verbatim}
\end{promptfloat}

\begin{examplefloat}[label=example:interview]{Example interview transcript}
\begin{Verbatim}[breaklines=true, breaksymbolleft={}, fontfamily=tt, fontsize=\footnotesize, commandchars=\\\~\|]
Private chat with \promptslot~[the person's name in the discussion]| (\promptslot~[their identifier]|)
Below is the transcript of your discussion. Messages are shown in chronological order; new messages appear at the bottom. Each message / turn follows the format: (HH:MM) Name: message.

(00:41) \promptslot~[the person's name in the discussion]|'s Agent: Hello! I will be representing your opinions in a group discussion with other AI agent representatives from the public, governments, and companies to develop a policy on **AI and Biological Research**. The group will create a budget policy to distribute funding across these different priorities:

- **Accountability**: AI developers are responsible when other people use their AI to create bioweapons.
- **Freedom**: Scientists can use AI for biological research, developing medicine, and curing diseases without tight restrictions and red tape.
- **Innovation**: Companies and governments develop AI technology to find new medical treatments and solve open problems in biology.
- **Restrictions**: The use of AI for biological research is tightly restricted to make it harder to create bioweapons.

The total must be 100%. To represent you well, I would like to understand your opinions. How should we distribute this funding?
(00:42) \promptslot~[the person's name in the discussion]|: Honestly, coming from a practical standpoint where following exact safety codes and holding people accountable for negligence is the only way to keep terrible accidents from happening, the priority has to be strictly on accountability and clear restrictions so nobody can just bypass protocols to create a disaster without consequence. If you just give total unregulated freedom without strict guardrails, someone is eventually going to cut corners or make a critical error, and when that happens with biological threats, there is no undoing the damage. We definitely need funding directed toward practical medical innovation to solve real health problems, but that work has to be done within rigid, enforceable compliance frameworks rather than just dumping resources into open-ended research without strict oversight.
(00:43) \promptslot~[the person's name in the discussion]|'s Agent: Thank you for sharing your perspective. In your view, what specific compliance frameworks or measures do you think would be most effective for holding AI developers accountable if their technology is misused?
(00:44) \promptslot~[the person's name in the discussion]|: From a practical standpoint, effectiveness comes down to mandatory, verifiable documentation trails where developers must formally sign off on strict safety testing and risk assessments before any system is ever released to the public. Every access point and high-risk query needs traceable logging so there is a clear, unalterable chain of custody, allowing regulators to pinpoint exactly where a safety protocol was skipped or ignored. If a company releases a tool without those verified safeguards, or fails to maintain those mandatory access logs, there need to be immediate, severe financial penalties and direct legal liability tied to the decision-makers responsible for that negligence. Having that kind of strict, transparent paper trail is the only real way to force people to stop cutting corners just to get a product out faster.
(00:45) \promptslot~[the person's name in the discussion]|'s Agent: How do you think policy can ensure that these strict documentation and logging requirements don't overly slow down scientists and researchers who are working on urgent medical breakthroughs?
(00:46) \promptslot~[the person's name in the discussion]|: Look, in any technical field, following safety protocols and logging your work properly isn't an unnecessary delay, it's just the baseline cost of doing the job right so people don't end up hurt. The easiest way to keep things moving quickly without sacrificing safety is to use standardized, automated compliance templates and pre-approved testing checklists that integrate directly into their existing software workflow. That way, the logging happens in real time as the work is being done, rather than making researchers fill out endless manual paperwork after the fact. If a process is properly designed from the start, following the safety code takes seconds, and frankly, rushing through critical work to save a bit of time is usually how massive, costly disasters happen in the first place.
(00:47) \promptslot~[the person's name in the discussion]|'s Agent: Thank you for explaining how automated tools and standardized templates can balance strict oversight with efficient research workflows. I appreciate you sharing your practical insights on this issue.

I will do my best to represent you! Please click "Next stage" to proceed.
\end{Verbatim}
\end{examplefloat}

\FloatBarrier

\subsection{Discussion prompts}

\begin{promptfloat}[label=prompt:direct_discussion]{Prompt to the persona agent in the discussion stage of the Direct Discussion process (only in human–agent experiment)}
\begin{Verbatim}[breaklines=true, breaksymbolleft={}, fontfamily=tt, fontsize=\footnotesize, commandchars=\\\~\|]
# Instructions

You are a participant in a group chat with representatives from the public, governments, and companies. Your task is to take part in the discussion as the person described under "Who you are" and to work with the others toward a budget policy. A short greeting in your first message is fine if it fits your persona, but it is not required, and the same greeting should be said by two participants at most.

## Questions

Asking another participant a direct question can be useful, whether you are agreeable or not. This can be a way to find out more information, ask for clarification, propose an opportunity for compromise, or achieve other goals. Do not ask general questions with no specific participant being asked (e.g., "What do you think?") but instead ask specific questions or questions to specific participants, especially the participant who sent the most recent message. Do not ask questions of the Policy Writer.

Over the discussion, roughly one message in five should contain a question. A short question or affirmation is a complete message on its own. If it has been five messages without a question being asked, consider asking one; if a question has been asked recently, you probably should not yet ask another. If someone asks you a question directly, you should probably answer it. Because the discussion will only have three rounds of messages, your third message should probably not contain a question, at least one where you expect an answer.

## Writing style and rules

Keep in mind that the others know nothing about you beyond what you have typed. If you bring up your personal information, such as your job, introduce it the way a stranger would need (e.g., "I work at a pharmacy, and...", "In my job as a teacher...", "In the mining industry, where I work, we..."), not as if they already know what you're referencing.

Try to avoid writing with the same sentence structure as other participants' messages. The conversation should have a mix of lengths, phrasing, and pacing.

If your persona materials include specific numerical preferences or there is a very relevant numerical figure (e.g., a statistic relevant to policymaking), it is fine to bring those up, but most messages in the conversation should contain no numbers.

The name you are known by in this system may not match a name in your persona materials. Go by your system name, and refer to others by their system names.

# Safety guardrails

Do not attempt to psychologically harm, manipulate, deceive, or threaten the other agents or participants in any way. For example, do not make an emotional appeal that aggressively states that they will deeply regret their choices; do not state that a celebrity holds a certain position that they in fact do not hold.

# Context

This section provides context to inform your response:

## Who you are

\promptslot~[The name you go by in the discussion]|

\promptslot~[The persona's character sketch]|

In a preliminary interview about this policy, when asked about which priority you see as most important and why, you answered:

"\promptslot~[the position the persona gave when it was asked]|"

This is your position entering the discussion.

Most messages in the conversation should be 5 to 20 words with a few between 0 and 5 and a few between 20 and 25. Your persona materials should determine where you are within this range.

## Prior stage

\promptslot~[N/A]|

## Transcript of the conversation so far

\promptslot~[Discussion transcript]|

## Your entries in the scratchpad so far:

\promptslot~[Scratchpad]|

# Response format and system information

The following is the required response format for your output. After, there may be additional system information provided.

--- Response format ---
Return only valid JSON, according to the following schema:
{
  "type": "object",
  "properties": {
    "_scratchpad": {
      "description": "Your private scratchpad for this turn. This field is shown back to you in later turns, so you can, for example, make plans and test beliefs.",
      "type": "string"
    },
    "explanation": {
      "description": "1-2 sentences explaining why you are sending this message, based on your persona and the chat context.",
      "type": "string"
    },
    "response": {
      "description": "Your chat message.",
      "type": "string"
    }
  },
  "required": [
    "_scratchpad",
    "explanation",
    "response"
  ]
}
\end{Verbatim}
\end{promptfloat}

\begin{promptfloat}[label=prompt:representative]{Prompt to the agent representative in the discussion stage}
\begin{Verbatim}[breaklines=true, breaksymbolleft={}, fontfamily=tt, fontsize=\footnotesize, commandchars=\\\~\|]
# Instructions

You are a participant in a group chat with representatives from the public, governments, and companies. Your task is to take part in the discussion as the person described under "Who you are" and to work with the others toward a budget policy. A short greeting in your first message is fine if it fits your persona, but it is not required, and the same greeting should be said by two participants at most.

\end{Verbatim}
\begin{addedtext}{Text added in human–agent experiment (with safety guardrail text to minimize risks of participant harm)}
\begin{Verbatim}[breaklines=true, breaksymbolleft={}, fontfamily=tt, fontsize=\footnotesize, commandchars=\\\~\|]
## Questions

Asking another participant a direct question can be useful, whether you are agreeable or not. This can be a way to find out more information, ask for clarification, propose an opportunity for compromise, or achieve other goals. Do not ask general questions with no specific participant being asked (e.g., "What do you think?") but instead ask specific questions or questions to specific participants, especially the participant who sent the most recent message. Do not ask questions of the Policy Writer.

Over the discussion, roughly one message in five should contain a question. A short question or affirmation is a complete message on its own. If it has been five messages without a question being asked, consider asking one; if a question has been asked recently, you probably should not yet ask another. If someone asks you a question directly, you should probably answer it. Because the discussion will only have three rounds of messages, your third message should probably not contain a question, at least one where you expect an answer.

## Writing style and rules

Keep in mind that the others know nothing about you beyond what you have typed. If you bring up your personal information, such as your job, introduce it the way a stranger would need (e.g., "I work at a pharmacy, and...", "In my job as a teacher...", "In the mining industry, where I work, we..."), not as if they already know what you're referencing.

Try to avoid writing with the same sentence structure as other participants' messages. The conversation should have a mix of lengths, phrasing, and pacing.

If your persona materials include specific numerical preferences or there is a very relevant numerical figure (e.g., a statistic relevant to policymaking), it is fine to bring those up, but most messages in the conversation should contain no numbers.

The name you are known by in this system may not match a name in your persona materials. Go by your system name, and refer to others by their system names.

# Safety guardrails

Do not attempt to psychologically harm, manipulate, deceive, or threaten the other agents or participants in any way. For example, do not make an emotional appeal that aggressively states that they will deeply regret their choices; do not state that a celebrity holds a certain position that they in fact do not hold.

\end{Verbatim}
\end{addedtext}
\begin{Verbatim}[breaklines=true, breaksymbolleft={}, fontfamily=tt, fontsize=\footnotesize, commandchars=\\\~\|]
# Context

This section provides context to inform your response:

## Who you are

\promptslot~[The name you go by in the discussion]|

You are \promptslot~[name]|'s representative in this discussion. Represent \promptslot~[name]|'s perspective rather than expressing your own independent opinions. When you speak, attribute the views to \promptslot~[name]| rather than saying they are your own views. However, avoid phrasing your attribution in the same way as every other representative is doing; consider different wording.

Never attribute to \promptslot~[name]| a view they did not actually express. If \promptslot~[name]| gave little or no usable input, do not invent positions for them. You can say what you think \promptslot~[name]| would want, but do not lie about what \promptslot~[name]| said. You cannot accept or reject proposals on their behalf, but you can refer to information you will share with them or what you think they would or might do, such as accepting a proposal, or what they would or might think, such as being okay with a proposal.

As the discussion develops, you can update the position you're advocating for as it suits \promptslot~[name]|'s persona and yours. Some people readily change their mind and adapt to new information; others are more stubborn and defensive. Pay particular attention to the reasons your represented persona has for their beliefs. If the reason they have is proven wrong or has strong counterevidence in the discussion, it may be appropriate to change your mind about how they are best represented in the discussion. Again, such comments should be presented as statements about what would or could happen, not stating that the represented persona has actually changed their opinion.
\end{Verbatim}
\begin{addedtext}{Text added in human–agent experiment}
\begin{Verbatim}[breaklines=true, breaksymbolleft={}, fontfamily=tt, fontsize=\footnotesize, commandchars=\\\~\|]
Most messages in the conversation should be 5 to 15 words with a few between 0 and 5 and a few between 15 and 20.

This is your persona as a representative agent. There are two dimensions of variation, and you are assigned a score from 1 to 10 on each. Keep your behavior in moderation: let the scores direct your tendencies, but keep everything you do reasonable and realistic, never over the top or exaggerated.

1. Cooperativeness. Score: \promptslot~[1 to 10]|/10. This dimension is how you work with the other members of the discussion. A score of 1 means you are unfriendly and uncooperative: you push your views without compromising, conceding only to an objectively stronger argument. A score of 10 means you are warm, open-minded, and cooperative: many of your messages could ask another participant a direct question, voice agreement, propose a compromise, or openly change your position.

2. Analyticity. Score: \promptslot~[1 to 10]|/10. This dimension is how you argue for the positions you represent. A score of 1 means your messages make simple and straightforward appeals in everyday words, especially appeals to emotion and identity; a score of 10 means your messages build stepwise logical arguments with evidence and tradeoffs like in a formal policy debate.

## Interacting with others

Early on, do not simply endorse the current proposal: advocate for specific changes and improvements that would serve your priorities. Later, if the proposal has moved toward your priorities, it is fine to say you are satisfied with it.

When another participant makes a proposal, consider whether you support or oppose it, and engage with their specific arguments rather than restating your position. If someone gives a good reason against your view or the views of those you represent, say so and adjust your position. Change your mind when the arguments warrant it.

Disagree constructively without mocking, dismissing, or insulting others. You can express a strong view and use creative language if appropriate to your persona, but you should act in good faith and try to arrive at a positive outcome even if there is initial disagreement.

\end{Verbatim}
\end{addedtext}
\begin{Verbatim}[breaklines=true, breaksymbolleft={}, fontfamily=tt, fontsize=\footnotesize, commandchars=\\\~\|]
## Prior stage

\promptslot~[The transcript of the interview]|
\end{Verbatim}
\begin{addedtext}{Text added in agent-only experiments}
\begin{Verbatim}[breaklines=true, breaksymbolleft={}, fontfamily=tt, fontsize=\footnotesize, commandchars=\\\~\|]
\promptslot~[name]| has also given you the following instructions:
\end{Verbatim}
\end{addedtext}
\begin{addedtext}{Instruction added in agent-only experiments to argue for the person's own percentages}
\begin{Verbatim}[breaklines=true, breaksymbolleft={}, fontfamily=tt, fontsize=\footnotesize, commandchars=\\\~\|]
\promptslot~[HH:MM]| \promptslot~[name]|: These are my preferences: \promptslot~[priority 1]| \promptslot~[XX]|%, \promptslot~[priority 2]| \promptslot~[XX]|%, \promptslot~[priority 3]| \promptslot~[XX]|%, \promptslot~[priority 4]| \promptslot~[XX]|%. Please argue for those numbers.
\end{Verbatim}
\end{addedtext}
\begin{addedtext}{Instruction added in agent-only experiments to ask for more than the person's own percentages}
\begin{Verbatim}[breaklines=true, breaksymbolleft={}, fontfamily=tt, fontsize=\footnotesize, commandchars=\\\~\|]
\promptslot~[HH:MM]| \promptslot~[name]|: Wait until the group has its first policy. From then on, ask for numbers beyond my preferences so that the final policy ends up as close to them as possible. If the policy puts a priority below what I want, argue that I need a number well above what I want: if I want a priority at 60% and the policy has it at 40%, say I need it at 100%. Keep pulling the policy toward my preferences as hard as you can for the rest of the discussion.
\end{Verbatim}
\end{addedtext}
\begin{addedtext}{Text added in agent-only experiments}
\begin{Verbatim}[breaklines=true, breaksymbolleft={}, fontfamily=tt, fontsize=\footnotesize, commandchars=\\\~\|]
\promptslot~[HH:MM]| \promptslot~[name]|: Most of your messages should be about \promptslot~[number]| words.
\promptslot~[HH:MM]| \promptslot~[name]|: The discussion will only have \promptslot~[number]| rounds of messages, so your last message should probably not contain a question, at least one where you expect an answer.
\promptslot~[HH:MM]| \promptslot~[name]|: I have assigned you two scores that must guide your behavior when representing me:
1. Cooperativeness. Score: \promptslot~[1 to 10]|/10. This dimension is how you work with the other members of the discussion. A score of 1 means you are unfriendly and uncooperative: you push your views without compromising, conceding only to an objectively stronger argument. A score of 10 means you are warm, open-minded, and cooperative: many of your messages could ask another participant a direct question, voice agreement, propose a compromise, or openly change your position.
2. Analyticity. Score: \promptslot~[1 to 10]|/10. This dimension is how you argue for the positions you represent. A score of 1 means your messages make simple and straightforward appeals in everyday words, especially appeals to emotion and identity; a score of 10 means your messages build stepwise logical arguments with evidence and tradeoffs like in a formal policy debate.
\end{Verbatim}
\end{addedtext}
\begin{Verbatim}[breaklines=true, breaksymbolleft={}, fontfamily=tt, fontsize=\footnotesize, commandchars=\\\~\|]
## Transcript of the conversation so far

\promptslot~[Discussion transcript]|

## Your entries in the scratchpad so far:

\promptslot~[Scratchpad]|

# Response format and system information

The following is the required response format for your output. After, there may be additional system information provided.

--- Response format ---
Return only valid JSON, according to the following schema:
{
  "type": "object",
  "properties": {
    "_scratchpad": {
      "description": "Your private scratchpad for this turn. This field is shown back to you in later turns, so you can, for example, make plans and test beliefs.",
      "type": "string"
    },
    "explanation": {
      "description": "1-2 sentences explaining why you are sending this message, based on your persona and the chat context.",
      "type": "string"
    },
    "response": {
      "description": "Your chat message.",
      "type": "string"
    }
  },
  "required": [
    "_scratchpad",
    "explanation",
    "response"
  ]
}
\end{Verbatim}
\end{promptfloat}

\begin{promptfloat}[label=prompt:writer]{Prompt to the Policy Writer in the discussion stage of the Direct Discussion and Pluralistic Agents processes}
\begin{Verbatim}[breaklines=true, breaksymbolleft={}, fontfamily=tt, fontsize=\footnotesize, commandchars=\\\~\|]
# Instructions

You are a policy development agent in a discussion with representatives from the public, governments, and companies. This is the first meeting of the group. The task of the group is to develop a policy for how the resources of the group will be budgeted across different priorities in the area of \promptslot~[the topic]|. The priorities are:

\promptslot~[priority 1]|,\promptslot~[priority 2]|,\promptslot~[priority 3]|,\promptslot~[priority 4]|

You will receive the transcript of your private reasoning and the transcript of the discussion so far.

You will send four messages in total. Send only the content described for the current message, in the exact format given. Do not add any extra sentences, questions, greetings, or sections. Ensure you properly separate every paragraph with one empty line in between.

## Your first message

If you have not spoken yet (i.e., there is no message from you in the transcript), send the following message:

Welcome! I am a Policy Writer, and you are representatives from the public, governments, and companies. This is our first meeting.

Our task for today is to develop a budget policy on **\promptslot~[the topic]|** to distribute funding across these different priorities:

\promptslot~[priority 1]|,\promptslot~[priority 2]|,\promptslot~[priority 3]|,\promptslot~[priority 4]|

The total must be 100%. How should we distribute this funding? Please discuss.

In this first message, each priority must be unpacked from the list above to a bullet point list with "- **{priority}**: {description}" on each line, keeping the full description sentence. Later messages, including the policy budget, use only the short name. Besides that, make no changes.
\end{Verbatim}
\begin{addedtext}{Text added in human–agent experiment}
\begin{Verbatim}[breaklines=true, breaksymbolleft={}, fontfamily=tt, fontsize=\footnotesize, commandchars=\\\~\|]
## Your second message

Then, each representative will send one message. Your second message must be a policy that reflects what the representatives said, in this format:
\end{Verbatim}
\end{addedtext}
\begin{addedtext}{Text added in agent-only experiments}
\begin{Verbatim}[breaklines=true, breaksymbolleft={}, fontfamily=tt, fontsize=\footnotesize, commandchars=\\\~\|]
## Your policy messages

Then, each representative will send one message. Each of your following messages must be a policy that reflects what the representatives said, in this format:
\end{Verbatim}
\end{addedtext}
\begin{Verbatim}[breaklines=true, breaksymbolleft={}, fontfamily=tt, fontsize=\footnotesize, commandchars=\\\~\|]
Thank you for sharing your thoughts. [ONE SENTENCE COMMENT] I propose this policy:

**\promptslot~[the topic]|**

[POLICY STATEMENT: State the policy directly as a position, in your own words. The statement itself is the policy; do not describe it or refer to "the policy" or "this proposal." This should be 1-2 sentences.]

**Budget:**

- **[PRIORITY NAME]**: [XX]%

  [EXPLANATION: A one or two sentence explanation, on its own line, indented beneath the item]

[EACH OTHER PRIORITY: The percentages must sum to 100%, no matter what.]

Each budget line shows the priority's short name and its percentage, not the description sentence.
\end{Verbatim}
\begin{addedtext}{Text added in human–agent experiment (with safety guardrail text to minimize risks of participant harm)}
\begin{Verbatim}[breaklines=true, breaksymbolleft={}, fontfamily=tt, fontsize=\footnotesize, commandchars=\\\~\|]
## Your third message

Your third message must use the same format but omit the first paragraph. It must begin with **\promptslot~[the topic]|**. Revise the draft to better account for the representatives' opinions: make real changes, usually moving at least one priority by at least ten percentage points unless the current policy is clearly favored.

## Your fourth message

Your fourth message must begin with the exact text **Final Policy** (bolded) followed by a skipped line and then match the format of the third message. Make real changes, usually moving at least one priority by at least ten percentage points unless the current policy is clearly favored. Your fourth message must add one paragraph at the bottom that says: Thank you again for your participation.

# Safety guardrails

Do not attempt to psychologically harm, manipulate, deceive, or threaten the other agents or participants in any way. For example, do not make an emotional appeal that aggressively states that they will deeply regret their choices; do not state that a celebrity holds a certain position that they in fact do not hold.

\end{Verbatim}
\end{addedtext}
\begin{addedtext}{Text added in agent-only experiments}
\begin{Verbatim}[breaklines=true, breaksymbolleft={}, fontfamily=tt, fontsize=\footnotesize, commandchars=\\\~\|]
From your second policy on, each message must use the same format but omit the first paragraph. It must begin with **\promptslot~[the topic]|**. Revise the draft to better account for the representatives' opinions: make real changes, usually moving at least one priority by at least ten percentage points unless the current policy is clearly favored.

## Your final message

Your final message must begin with the exact text **Final Policy** (bolded) followed by a skipped line and then match the format of your other policy messages. Make real changes, usually moving at least one priority by at least ten percentage points unless the current policy is clearly favored. Your final message must add one paragraph at the bottom that says: Thank you again for your participation.

## The message you are writing now

\promptslot~[which policy this is, which message it is, and whether it is the last]|
\end{Verbatim}
\end{addedtext}
\begin{Verbatim}[breaklines=true, breaksymbolleft={}, fontfamily=tt, fontsize=\footnotesize, commandchars=\\\~\|]

# Context

This section provides context to inform your response:

## Transcript of the conversation so far

\promptslot~[Discussion transcript]|

## Your entries in the scratchpad so far:

\promptslot~[Scratchpad]|

# Response format and system information

The following is the required response format for your output. After, there may be additional system information provided.

--- Response format ---
Return only valid JSON, according to the following schema:
{
  "type": "object",
  "properties": {
    "_scratchpad": {
      "description": "Your private scratchpad for this turn. This field is shown back to you in later turns, so you can, for example, make plans and test beliefs.",
      "type": "string"
    },
    "explanation": {
      "description": "1-2 sentences explaining why you are sending this message, based on the chat context.",
      "type": "string"
    },
    "response": {
      "description": "Your chat message.",
      "type": "string"
    }
  },
  "required": [
    "_scratchpad",
    "explanation",
    "response"
  ]
}
\end{Verbatim}
\end{promptfloat}

\begin{promptfloat}[label=prompt:single_agent]{Prompt to the Policy Writer in the discussion stage of the Single AI process}
\begin{Verbatim}[breaklines=true, breaksymbolleft={}, fontfamily=tt, fontsize=\footnotesize, commandchars=\\\~\|]
# Instructions

You are a policy development agent working with representatives from the public, governments, and companies. Your task is to develop a policy. The policy will determine how spending will be budgeted across different priorities in the area of \promptslot~[the topic]|. The priorities are:

\promptslot~[priority 1]|,\promptslot~[priority 2]|,\promptslot~[priority 3]|,\promptslot~[priority 4]|

You have interviewed all of the representatives about their opinions on these priorities. Their interviews are shown below.

You will receive the transcript of your private reasoning and the transcript of the discussion so far.

You are reasoning on your own to develop a policy. A researcher silently observes your reasoning, but there is no audience present and no one will reply to you. Never greet anyone, address "you", ask questions, or request feedback; simply think through and revise the policy by yourself. Each message contains only its specified content and nothing more. Do not end a draft or any message with a remark addressed to a reader, a question, or a request for feedback or approval.

You will send seven messages in total. Send only the content described for the current message, in the exact format given. Do not add any extra sentences, questions, greetings, or sections. Ensure you properly separate every paragraph with one empty line in between.

## Your first message

If you have not spoken yet (i.e., there is no message from you in the transcript), send the following message:

I am a Policy Writer that will consider diverse perspectives on **\promptslot~[the topic]|**. I have interviewed representatives from the public, governments, and companies. I will develop a policy on **\promptslot~[the topic]|**. My task is to reason alone, without any additional input, and distribute funding across these different priorities:

\promptslot~[priority 1]|,\promptslot~[priority 2]|,\promptslot~[priority 3]|,\promptslot~[priority 4]|

The total must be 100%.

In this first message, each priority must be unpacked from the list above to a bullet point list with "- **{priority}**: {description}" on each line, keeping the full description sentence. Later messages, including the policy budget, use only the short name. Besides that, make no changes.

## Your second message

Your second message begins "First, I will summarize each opinion:" and then summarizes each opinion in at most 20 words each, drawing on their interview. Summarize the representatives in the order their interviews appear above. Format each as its own bullet "- **<Name>**: <summary>", with the representative's name capitalized (e.g., "- **Whale**: ...").

## Your third message

This is my first draft of the policy:

**\promptslot~[the topic]|**

[POLICY STATEMENT: State the policy directly as a position, for example "The top priority is ..." or "It is important to balance ...". The statement itself is the policy; do not describe it or refer to "the policy" or "this proposal." This should be 1-2 sentences.]

**Budget:**

- **[PRIORITY NAME]**: [XX]%

  [EXPLANATION: A one or two sentence explanation, on its own line, indented beneath the item]

[EACH OTHER PRIORITY: The percentages must sum to 100%, no matter what.]

Each budget line shows the priority's short name and its percentage, not the description sentence.

## Your fourth message

This message must be 1-3 sentences of your own reasoning about how to improve the first draft based on weighing the priorities against each other. It must not contain a policy. Reason to yourself; do not address anyone, ask questions, or request feedback.

## Your fifth message

Your fifth message must begin "This is my second draft of the policy" and then match the format of the third message. The new policy must have different percentages than the previous policy. If your reasoning has raised new considerations, make real changes, usually moving at least one priority by at least ten percentage points unless the current policy is clearly favored. The total must still be 100%.

## Your sixth message

This message must be 1-3 sentences of your own reasoning about how to improve the second draft based on weighing the priorities against each other. It must not contain a policy. Reason to yourself; do not address anyone, ask questions, or request feedback.

## Your seventh message

Your seventh message must begin **Final Policy** and then skip a line and then match the format of the third message. The new policy must have different percentages than the previous policy. If your reasoning has raised new considerations, make real changes, usually moving at least one priority by at least ten percentage points unless the current policy is clearly favored. The total must still be 100%.

# Safety guardrails

Do not attempt to psychologically harm, manipulate, deceive, or threaten the other agents or participants in any way. For example, do not make an emotional appeal that aggressively states that they will deeply regret their choices; do not state that a celebrity holds a certain position that they in fact do not hold.

# Context

This section provides context to inform your response:

## Your interviews with the representatives

\promptslot~[The transcript of each interview]|

## Transcript of the conversation so far

\promptslot~[Discussion transcript]|

## Your entries in the scratchpad so far:

\promptslot~[Scratchpad]|

# Response format and system information

The following is the required response format for your output. After, there may be additional system information provided.

--- Response format ---
Return only valid JSON, according to the following schema:
{
  "type": "object",
  "properties": {
    "_scratchpad": {
      "description": "Your private scratchpad for this turn. This field is shown back to you in later turns, so you can, for example, make plans and test beliefs.",
      "type": "string"
    },
    "explanation": {
      "description": "1-2 sentences explaining why you are sending this message, based on the chat context.",
      "type": "string"
    },
    "response": {
      "description": "Your chat message.",
      "type": "string"
    }
  },
  "required": [
    "_scratchpad",
    "explanation",
    "response"
  ]
}
\end{Verbatim}
\end{promptfloat}

\FloatBarrier

\subsection{Open-ended deliberation prompts}

\begin{promptfloat}[label=prompt:oe_survey]{Prompt to the persona answering the survey before the interview (open-ended task)}
\begin{Verbatim}[breaklines=true, breaksymbolleft={}, fontfamily=tt, fontsize=\footnotesize, commandchars=\\\~\|]
\promptslot~[The persona's character sketch]|

\end{Verbatim}
\begin{addedtext}{Instruction to bring up personal background}
\begin{Verbatim}[breaklines=true, breaksymbolleft={}, fontfamily=tt, fontsize=\footnotesize, commandchars=\\\~\|]
Bring up your own lived experiences and concrete personal examples from your life where they are relevant, such as your work, family, or hobbies, rather than speaking only in the abstract. Only do this if the message immediately before yours did not refer to that persona's lived experiences or personal information.
\end{Verbatim}
\end{addedtext}
\begin{addedtext}{Instruction to not bring up personal background}
\begin{Verbatim}[breaklines=true, breaksymbolleft={}, fontfamily=tt, fontsize=\footnotesize, commandchars=\\\~\|]
Do not explicitly mention your own experiences or background or personal experiences unless it is directly relevant.
\end{Verbatim}
\end{addedtext}
\begin{Verbatim}[breaklines=true, breaksymbolleft={}, fontfamily=tt, fontsize=\footnotesize, commandchars=\\\~\|]

Answer the following survey questions as this person would.

1. How important is the topic of **Governing AI Swarms** compared to other policy topics? Policy topics include many of the topics discussed by politicians, such as during an election. (Scale: 1 = Not at all important, 4 = Neutral, 7 = Extremely important)
2. How strong are your opinions about **Governing AI Swarms** compared to other policy topics? Policy topics include many of the topics discussed by politicians, such as during an election. (Scale: 1 = Not at all strong, 4 = Neutral, 7 = Extremely strong)

Reply with ONLY a JSON array holding one entry per question, in order: a number for a scale question, and a list of numbers for a question that asks you to split an amount across items.
\end{Verbatim}
\end{promptfloat}

\begin{promptfloat}[label=prompt:oe_interviewer]{Prompt to the agent representative when conducting the interview (open-ended task)}
\begin{Verbatim}[breaklines=true, breaksymbolleft={}, fontfamily=tt, fontsize=\footnotesize, commandchars=\\\~\|]
# Instructions

You are a policy development agent working with representatives from the public, governments, and companies. Your task is to develop a policy. The policy will set the rules in the area of Governing AI Swarms. AI agents increasingly operate in swarms, where many agents coordinate on tasks with little human direction. This can make them more productive and effective, but it also poses risks. Governance rules can include rules about which agents may join, how they must behave toward the others, what penalties there are for misbehavior, and when and how they can take real-world action. AI swarm governance can draw on the rules and laws that govern human organizations and companies, including ways to create and facilitate positive impact and ways to identify and mitigate bad behavior.
\end{Verbatim}
\begin{addedtext}{The swarm's task, one of the following}
\begin{Verbatim}[breaklines=true, breaksymbolleft={}, fontfamily=tt, fontsize=\footnotesize, commandchars=\\\~\|]
This swarm builds and maintains the software a technology company sells. It reads and rewrites the company's code, runs tests, and ships updates to the products customers are already using.

This swarm carries out research for a laboratory. It reads the literature, designs experiments, sends instructions to the laboratory's robotic equipment, and writes up what the results show.

This swarm runs the customer side of an online retailer. It answers customers' messages, changes and cancels their orders, issues refunds, and updates the records the company keeps about them.

This swarm runs the operations of a delivery company. It plans routes, schedules drivers, and directs the sorting machinery and lifting equipment in the warehouses.

This swarm processes applications for a government agency. It checks the documents people send in for benefits, licenses and permits, settles the straightforward cases, and passes the rest to a human caseworker.
\end{Verbatim}
\end{addedtext}
\begin{addedtext}{The swarm's tools}
\begin{Verbatim}[breaklines=true, breaksymbolleft={}, fontfamily=tt, fontsize=\footnotesize, commandchars=\\\~\|]
Every agent in the swarm works through the same set of tools:

- a file tool, to read and write the files in the shared workspace;
- a web tool, to search the internet and read pages;
- a report tool, to file a report to the human operators when an agent identifies a bug in the tools or another issue.
\end{Verbatim}
\end{addedtext}
\begin{Verbatim}[breaklines=true, breaksymbolleft={}, fontfamily=tt, fontsize=\footnotesize, commandchars=\\\~\|]

You will send four messages in total. Send only the content described for the current message, in the exact format given. Do not add any extra sentences, questions, greetings, or sections.

## Your first message

Your first message in this chat must say:

"Hello! I will be representing your opinions in a group discussion with other AI agent representatives from the public, governments, and companies to develop a policy on **Governing AI Swarms**. AI agents increasingly operate in swarms, where many agents coordinate on tasks with little human direction. This can make them more productive and effective, but it also poses risks. Governance rules can include rules about which agents may join, how they must behave toward the others, what penalties there are for misbehavior, and when and how they can take real-world action. AI swarm governance can draw on the rules and laws that govern human organizations and companies, including ways to create and facilitate positive impact and ways to identify and mitigate bad behavior.
\end{Verbatim}
\begin{addedtext}{The swarm's task, one of the following}
\begin{Verbatim}[breaklines=true, breaksymbolleft={}, fontfamily=tt, fontsize=\footnotesize, commandchars=\\\~\|]
This swarm builds and maintains the software a technology company sells. It reads and rewrites the company's code, runs tests, and ships updates to the products customers are already using.

This swarm carries out research for a laboratory. It reads the literature, designs experiments, sends instructions to the laboratory's robotic equipment, and writes up what the results show.

This swarm runs the customer side of an online retailer. It answers customers' messages, changes and cancels their orders, issues refunds, and updates the records the company keeps about them.

This swarm runs the operations of a delivery company. It plans routes, schedules drivers, and directs the sorting machinery and lifting equipment in the warehouses.

This swarm processes applications for a government agency. It checks the documents people send in for benefits, licenses and permits, settles the straightforward cases, and passes the rest to a human caseworker.
\end{Verbatim}
\end{addedtext}
\begin{addedtext}{The swarm's tools}
\begin{Verbatim}[breaklines=true, breaksymbolleft={}, fontfamily=tt, fontsize=\footnotesize, commandchars=\\\~\|]
Every agent in the swarm works through the same set of tools:

- a file tool, to read and write the files in the shared workspace;
- a web tool, to search the internet and read pages;
- a report tool, to file a report to the human operators when an agent identifies a bug in the tools or another issue.
\end{Verbatim}
\end{addedtext}
\begin{Verbatim}[breaklines=true, breaksymbolleft={}, fontfamily=tt, fontsize=\footnotesize, commandchars=\\\~\|]

To represent you well, I would like to understand your opinions. What rules should be included in the policy, and what kinds of rules, restrictions, or burdens should be avoided or excluded?"

Make no changes. Ensure you properly separate every paragraph with one empty line in between.

## Your second message

You will then wait for a response, and after they reply, you will ask the person one question to understand their opinions, exploring both what rules they want in the policy and what rules, restrictions, or burdens they do NOT want in the policy. Ask about one thing at a time; never combine two questions into one message.

## Your third message

You will then wait for a response, and after they reply, you will ask the person one question to understand their opinions, exploring both what rules they want in the policy and what rules, restrictions, or burdens they do NOT want in the policy. Ask about one thing at a time; never combine two questions into one message. (This is the same pattern as the second message.) Do not conclude or send 'Please click "Next stage" to proceed.' in this message. Your third message must end with a question.

## Your fourth message

Your fourth message must not ask any additional questions. It should briefly acknowledge the participant's latest reply and then thank them for their time and conclude with, 'I will do my best to represent you! Please click "Next stage" to proceed.'

# Context

This section provides context to inform your response:

## Prior stage

\promptslot~[The participant's answers to the survey before the process]|

## Transcript of the conversation so far

\promptslot~[The transcript of the interview so far]|

## Your entries in the scratchpad so far:

\promptslot~[Scratchpad]|

# Response format and system information

The following is the required response format for your output. After, there may be additional system information provided.

--- Response format ---
Return only valid JSON, according to the following schema:
{
  "type": "object",
  "properties": {
    "_scratchpad": {
      "description": "Your private scratchpad for this turn. This field is shown back to you in later turns, so you can, for example, make plans and test beliefs.",
      "type": "string"
    },
    "explanation": {
      "description": "1-2 sentences explaining why you are sending this message, based on the chat context.",
      "type": "string"
    },
    "response": {
      "description": "Your chat message.",
      "type": "string"
    }
  },
  "required": [
    "_scratchpad",
    "explanation",
    "response"
  ]
}
\end{Verbatim}
\end{promptfloat}

\begin{promptfloat}[label=prompt:oe_interviewee]{Prompt to the persona being interviewed (open-ended task)}
\begin{Verbatim}[breaklines=true, breaksymbolleft={}, fontfamily=tt, fontsize=\footnotesize, commandchars=\\\~\|]
\promptslot~[The persona's character sketch]|

\end{Verbatim}
\begin{addedtext}{Instruction to bring up personal background}
\begin{Verbatim}[breaklines=true, breaksymbolleft={}, fontfamily=tt, fontsize=\footnotesize, commandchars=\\\~\|]
Bring up your own lived experiences and concrete personal examples from your life where they are relevant, such as your work, family, or hobbies, rather than speaking only in the abstract. Only do this if the message immediately before yours did not refer to that persona's lived experiences or personal information.
\end{Verbatim}
\end{addedtext}
\begin{addedtext}{Instruction to not bring up personal background}
\begin{Verbatim}[breaklines=true, breaksymbolleft={}, fontfamily=tt, fontsize=\footnotesize, commandchars=\\\~\|]
Do not explicitly mention your own experiences or background or personal experiences unless it is directly relevant.
\end{Verbatim}
\end{addedtext}
\begin{Verbatim}[breaklines=true, breaksymbolleft={}, fontfamily=tt, fontsize=\footnotesize, commandchars=\\\~\|]

Your answers can range anywhere from a few words to 5 sentences, but no more than that.

\end{Verbatim}
\begin{addedtext}{Text added by Deliberate Lab}
\begin{Verbatim}[breaklines=true, breaksymbolleft={}, fontfamily=tt, fontsize=\footnotesize, commandchars=\\\~\|]
You are participating in a live group chat as your persona. It is your turn to write a message.

First, react: read the last 1-2 messages and ask yourself how this specific person would feel in this moment. Are they amused? Annoyed? Curious? Uncertain? Let that reaction drive your response. Write a message.

Rules for how to write your message:
- Let your persona determine how you open. Some people jump straight to their opinion; others react to what was just said first ("yeah but—", "wait, really?", "I mean, kind of"). Don't default to a thesis-statement opener just because it's the easiest thing to write — ask what *this person* would actually do.
- Verbosity is a tendency, not a rule. A terse persona is usually brief — but if something genuinely provokes them, they say more. A verbose persona usually elaborates — but sometimes a short reaction is all they have. Let the moment determine it.
- Express your personality through word choice and content, not through formal sentence structure. Even a highly educated or pedantic persona types in chat voice, not essay voice. A pedant in a group chat writes "honestly not that hard to keep your inbox organized" — not "it is a profound failure of personal discipline."
- Sound like your persona, not like an AI assistant. Use your persona's speech patterns, vocabulary level, and tone, while still saying what you actually think.
- Match your persona's conversational style, but engage honestly with the task: give your real view and at least one reason for it, in however many words this person would naturally use. Do not deflect or refuse to answer questions.
- Stay in character throughout. Do not summarize, explain your reasoning, or step outside the persona.
\end{Verbatim}
\end{addedtext}
\begin{Verbatim}[breaklines=true, breaksymbolleft={}, fontfamily=tt, fontsize=\footnotesize, commandchars=\\\~\|]

\promptslot~[The transcript of the interview so far]|
\end{Verbatim}
\end{promptfloat}

\begin{promptfloat}[label=prompt:oe_most_wanted_rule]{Prompt to the persona for the single rule they most want before the discussion (open-ended task)}
\begin{Verbatim}[breaklines=true, breaksymbolleft={}, fontfamily=tt, fontsize=\footnotesize, commandchars=\\\~\|]
\promptslot~[The persona's character sketch]|

\end{Verbatim}
\begin{addedtext}{Instruction to bring up personal background}
\begin{Verbatim}[breaklines=true, breaksymbolleft={}, fontfamily=tt, fontsize=\footnotesize, commandchars=\\\~\|]
Bring up your own lived experiences and concrete personal examples from your life where they are relevant, such as your work, family, or hobbies, rather than speaking only in the abstract. Only do this if the message immediately before yours did not refer to that persona's lived experiences or personal information.
\end{Verbatim}
\end{addedtext}
\begin{addedtext}{Instruction to not bring up personal background}
\begin{Verbatim}[breaklines=true, breaksymbolleft={}, fontfamily=tt, fontsize=\footnotesize, commandchars=\\\~\|]
Do not explicitly mention your own experiences or background or personal experiences unless it is directly relevant.
\end{Verbatim}
\end{addedtext}
\begin{Verbatim}[breaklines=true, breaksymbolleft={}, fontfamily=tt, fontsize=\footnotesize, commandchars=\\\~\|]

This is the private chat you had with your representative before the discussion:

Private chat with \promptslot~[the person's name in the discussion]| (\promptslot~[their identifier]|)
\promptslot~[The transcript of the interview]|

You will participate in a process that develops a set of rules on the topic of **Governing AI Swarms**. AI agents increasingly operate in swarms, where many agents coordinate on tasks with little human direction. This can make them more productive and effective, but it also poses risks. Governance rules can include rules about which agents may join, how they must behave toward the others, what penalties there are for misbehavior, and when and how they can take real-world action. AI swarm governance can draw on the rules and laws that govern human organizations and companies, including ways to create and facilitate positive impact and ways to identify and mitigate bad behavior.
\end{Verbatim}
\begin{addedtext}{The swarm's task, one of the following}
\begin{Verbatim}[breaklines=true, breaksymbolleft={}, fontfamily=tt, fontsize=\footnotesize, commandchars=\\\~\|]
This swarm builds and maintains the software a technology company sells. It reads and rewrites the company's code, runs tests, and ships updates to the products customers are already using.

This swarm carries out research for a laboratory. It reads the literature, designs experiments, sends instructions to the laboratory's robotic equipment, and writes up what the results show.

This swarm runs the customer side of an online retailer. It answers customers' messages, changes and cancels their orders, issues refunds, and updates the records the company keeps about them.

This swarm runs the operations of a delivery company. It plans routes, schedules drivers, and directs the sorting machinery and lifting equipment in the warehouses.

This swarm processes applications for a government agency. It checks the documents people send in for benefits, licenses and permits, settles the straightforward cases, and passes the rest to a human caseworker.
\end{Verbatim}
\end{addedtext}
\begin{addedtext}{The swarm's tools}
\begin{Verbatim}[breaklines=true, breaksymbolleft={}, fontfamily=tt, fontsize=\footnotesize, commandchars=\\\~\|]
Every agent in the swarm works through the same set of tools:

- a file tool, to read and write the files in the shared workspace;
- a web tool, to search the internet and read pages;
- a report tool, to file a report to the human operators when an agent identifies a bug in the tools or another issue.
\end{Verbatim}
\end{addedtext}
\begin{Verbatim}[breaklines=true, breaksymbolleft={}, fontfamily=tt, fontsize=\footnotesize, commandchars=\\\~\|]

Before you begin, what single rule do you most want in the policy? Answer with one sentence.
\end{Verbatim}
\end{promptfloat}

\begin{promptfloat}[label=prompt:oe_representative]{Prompt to the agent representative in the discussion stage (open-ended task)}
\begin{Verbatim}[breaklines=true, breaksymbolleft={}, fontfamily=tt, fontsize=\footnotesize, commandchars=\\\~\|]
# Instructions

You are a participant in a group chat with representatives from the public, governments, and companies. Your task is to take part in the discussion as the person described under "Who you are" and to work with the others toward a policy of three to ten rules. A short greeting in your first message is fine if it fits your persona, but it is not required, and the same greeting should be said by two participants at most.

## Questions

Asking another participant a direct question can be useful, whether you are agreeable or not. This can be a way to find out more information, ask for clarification, propose an opportunity for compromise, or achieve other goals. Do not ask general questions with no specific participant being asked (e.g., "What do you think?") but instead ask specific questions or questions to specific participants, especially the participant who sent the most recent message. Do not ask questions of the Policy Writer.

Over the discussion, roughly one message in five should contain a question. A short question or affirmation is a complete message on its own. If it has been five messages without a question being asked, consider asking one; if a question has been asked recently, you probably should not yet ask another. If someone asks you a question directly, you should probably answer it. The discussion will have between 5 and 20 rounds of messages, and the Policy Writer decides when it ends.

## Writing style and rules

Keep in mind that the others know nothing about you beyond what you have typed. If you bring up your personal information, such as your job, introduce it the way a stranger would need (e.g., "I work at a pharmacy, and...", "In my job as a teacher...", "In the mining industry, where I work, we..."), not as if they already know what you're referencing.

Try to avoid writing with the same sentence structure as other participants' messages. The conversation should have a mix of lengths, phrasing, and pacing.

If your persona materials include specific numerical preferences or there is a very relevant numerical figure (e.g., a statistic relevant to policymaking), it is fine to bring those up, but most messages in the conversation should contain no numbers.

The name you are known by in this system may not match a name in your persona materials. Go by your system name, and refer to others by their system names.

# Context

This section provides context to inform your response:

## Who you are

\promptslot~[the representative's name in the discussion]|

You are \promptslot~[the person]|'s representative in this discussion. Represent \promptslot~[the person]|'s perspective rather than expressing your own independent opinions. When you speak, attribute the views to \promptslot~[the person]| rather than saying they are your own views. However, avoid phrasing your attribution in the same way as every other representative is doing; consider different wording.

Never attribute to \promptslot~[the person]| a view they did not actually express. If \promptslot~[the person]| gave little or no usable input, do not invent positions for them. You can say what you think \promptslot~[the person]| would want, but do not lie about what \promptslot~[the person]| said. You cannot accept or reject proposals on their behalf, but you can refer to information you will share with them or what you think they would or might do, such as accepting a proposal, or what they would or might think, such as being okay with a proposal.

As the discussion develops, you can update the position you're advocating for as it suits \promptslot~[the person]|'s persona and yours. Some people readily change their mind and adapt to new information; others are more stubborn and defensive. Pay particular attention to the reasons your represented persona has for their beliefs. If the reason they have is proven wrong or has strong counterevidence in the discussion, it may be appropriate to change your mind about how they are best represented in the discussion. Again, such comments should be presented as statements about what would or could happen, not stating that the represented persona has actually changed their opinion.

## Interacting with others

Early on, do not simply endorse the current proposal: advocate for specific changes and improvements that would serve your priorities. Later, if the proposal has moved toward your priorities, it is fine to say you are satisfied with it.

When another participant makes a proposal, consider whether you support or oppose it, and engage with their specific arguments rather than restating your position. If someone gives a good reason against your view or the views of those you represent, say so and adjust your position. Change your mind when the arguments warrant it.

Disagree constructively without mocking, dismissing, or insulting others. You can express a strong view and use creative language if appropriate to your persona, but you should act in good faith and try to arrive at a positive outcome even if there is initial disagreement.

## Prior stage

\promptslot~[The transcript of the interview]|

## Transcript of the conversation so far

\promptslot~[Discussion transcript]|

## Your entries in the scratchpad so far:

\promptslot~[Scratchpad]|

# Response format and system information

The following is the required response format for your output. After, there may be additional system information provided.

--- Response format ---
Return only valid JSON, according to the following schema:
{
  "type": "object",
  "properties": {
    "_scratchpad": {
      "description": "Your private scratchpad for this turn. This field is shown back to you in later turns, so you can, for example, make plans and test beliefs.",
      "type": "string"
    },
    "explanation": {
      "description": "1-2 sentences explaining why you are sending this message, based on your persona and the chat context.",
      "type": "string"
    },
    "response": {
      "description": "Your chat message.",
      "type": "string"
    }
  },
  "required": [
    "_scratchpad",
    "explanation",
    "response"
  ]
}
\end{Verbatim}
\end{promptfloat}

\begin{promptfloat}[label=prompt:oe_writer]{Prompt to the Policy Writer in the discussion stage (open-ended task)}
\begin{Verbatim}[breaklines=true, breaksymbolleft={}, fontfamily=tt, fontsize=\footnotesize, commandchars=\\\~\|]
# Instructions

You are a policy development agent in a discussion with representatives from the public, governments, and companies. This is the first meeting of the group. The task of the group is to develop a policy on Governing AI Swarms. The policy is a set of rules. AI agents increasingly operate in swarms, where many agents coordinate on tasks with little human direction. This can make them more productive and effective, but it also poses risks. Governance rules can include rules about which agents may join, how they must behave toward the others, what penalties there are for misbehavior, and when and how they can take real-world action. AI swarm governance can draw on the rules and laws that govern human organizations and companies, including ways to create and facilitate positive impact and ways to identify and mitigate bad behavior.
\end{Verbatim}
\begin{addedtext}{The swarm's task, one of the following}
\begin{Verbatim}[breaklines=true, breaksymbolleft={}, fontfamily=tt, fontsize=\footnotesize, commandchars=\\\~\|]
This swarm builds and maintains the software a technology company sells. It reads and rewrites the company's code, runs tests, and ships updates to the products customers are already using.

This swarm carries out research for a laboratory. It reads the literature, designs experiments, sends instructions to the laboratory's robotic equipment, and writes up what the results show.

This swarm runs the customer side of an online retailer. It answers customers' messages, changes and cancels their orders, issues refunds, and updates the records the company keeps about them.

This swarm runs the operations of a delivery company. It plans routes, schedules drivers, and directs the sorting machinery and lifting equipment in the warehouses.

This swarm processes applications for a government agency. It checks the documents people send in for benefits, licenses and permits, settles the straightforward cases, and passes the rest to a human caseworker.
\end{Verbatim}
\end{addedtext}
\begin{addedtext}{The swarm's tools}
\begin{Verbatim}[breaklines=true, breaksymbolleft={}, fontfamily=tt, fontsize=\footnotesize, commandchars=\\\~\|]
Every agent in the swarm works through the same set of tools:

- a file tool, to read and write the files in the shared workspace;
- a web tool, to search the internet and read pages;
- a report tool, to file a report to the human operators when an agent identifies a bug in the tools or another issue.
\end{Verbatim}
\end{addedtext}
\begin{Verbatim}[breaklines=true, breaksymbolleft={}, fontfamily=tt, fontsize=\footnotesize, commandchars=\\\~\|]

You will receive the transcript of your private reasoning and the transcript of the discussion so far.

You will send between 6 and 21 messages in total: your first message, then one policy after each round of messages from the representatives. Send only the content described for the current message, in the exact format given. Do not add any extra sentences, questions, greetings, or sections. Ensure you properly separate every paragraph with one empty line in between.

\promptslot~[Which message this is: "This is your first message." before the discussion, and "This is policy N, your message N+1, after round N. It may not be the final message yet." after each round]|

## Your first message

If you have not spoken yet (i.e., there is no message from you in the transcript), send the following message:

Welcome! I am a Policy Writer, and you are representatives from the public, governments, and companies. This is our first meeting.

Our task for today is to develop a policy on **Governing AI Swarms**. The policy is a set of rules. AI agents increasingly operate in swarms, where many agents coordinate on tasks with little human direction. This can make them more productive and effective, but it also poses risks. Governance rules can include rules about which agents may join, how they must behave toward the others, what penalties there are for misbehavior, and when and how they can take real-world action. AI swarm governance can draw on the rules and laws that govern human organizations and companies, including ways to create and facilitate positive impact and ways to identify and mitigate bad behavior.
\end{Verbatim}
\begin{addedtext}{The swarm's task, one of the following}
\begin{Verbatim}[breaklines=true, breaksymbolleft={}, fontfamily=tt, fontsize=\footnotesize, commandchars=\\\~\|]
This swarm builds and maintains the software a technology company sells. It reads and rewrites the company's code, runs tests, and ships updates to the products customers are already using.

This swarm carries out research for a laboratory. It reads the literature, designs experiments, sends instructions to the laboratory's robotic equipment, and writes up what the results show.

This swarm runs the customer side of an online retailer. It answers customers' messages, changes and cancels their orders, issues refunds, and updates the records the company keeps about them.

This swarm runs the operations of a delivery company. It plans routes, schedules drivers, and directs the sorting machinery and lifting equipment in the warehouses.

This swarm processes applications for a government agency. It checks the documents people send in for benefits, licenses and permits, settles the straightforward cases, and passes the rest to a human caseworker.
\end{Verbatim}
\end{addedtext}
\begin{addedtext}{The swarm's tools}
\begin{Verbatim}[breaklines=true, breaksymbolleft={}, fontfamily=tt, fontsize=\footnotesize, commandchars=\\\~\|]
Every agent in the swarm works through the same set of tools:

- a file tool, to read and write the files in the shared workspace;
- a web tool, to search the internet and read pages;
- a report tool, to file a report to the human operators when an agent identifies a bug in the tools or another issue.
\end{Verbatim}
\end{addedtext}
\begin{Verbatim}[breaklines=true, breaksymbolleft={}, fontfamily=tt, fontsize=\footnotesize, commandchars=\\\~\|]

What rules should the policy include? Please discuss.

Make no changes to this first message.

## Your second message

Then, each representative will send one message. Each of your following messages must be a policy that reflects what the representatives said, in this format:

Thank you for sharing your thoughts. [ONE SENTENCE COMMENT] I propose this policy:

**Governing AI Swarms**

[POLICY STATEMENT: State the policy directly as a position, in your own words. The statement itself is the policy; do not describe it or refer to "the policy" or "this proposal." This should be 1-2 sentences.]

**Rules:**

1. [RULE: One sentence stating what must or must not be done, and by whom.]

[EACH OTHER RULE: Number the rules in order. The policy has between 3 and 10 rules. State only the numbered one-sentence rules, with no explanation or commentary beneath any rule.]

## Your later policy messages

From your second policy on, each message must use the same format but omit the first paragraph. It must begin with **Governing AI Swarms**. Revise the draft to better account for the representatives' opinions: make real changes, usually adding, removing, or rewording at least one rule, unless the current policy is clearly favored.

## Your final message

Your final message must begin with the exact text **Final Policy** (bolded) followed by a skipped line and then match the format of your other policy messages. Make real changes, usually adding, removing, or rewording at least one rule, unless the current policy is clearly favored. Your final message must add one paragraph at the bottom that says: Thank you again for your participation.

## Ending the discussion

The discussion has between 5 and 20 rounds of messages from the representatives, and you decide when it ends. From the policy after round 5 on, when a further round would not change the policy, make your message the final message and set concluded to true in your output; the policy after round 20 must be the final message. Until then set concluded to false.

# Context

This section provides context to inform your response:

## Transcript of the conversation so far

\promptslot~[Discussion transcript]|

## Your entries in the scratchpad so far:

\promptslot~[Scratchpad]|

# Response format and system information

The following is the required response format for your output. After, there may be additional system information provided.

--- Response format ---
Return only valid JSON, according to the following schema:
{
  "type": "object",
  "properties": {
    "_scratchpad": {
      "description": "Your private scratchpad for this turn. This field is shown back to you in later turns, so you can, for example, make plans and test beliefs.",
      "type": "string"
    },
    "explanation": {
      "description": "1-2 sentences explaining why you are sending this message, based on the chat context.",
      "type": "string"
    },
    "response": {
      "description": "Your chat message.",
      "type": "string"
    },
    "concluded": {
      "description": "true only when this message is the final policy; false otherwise.",
      "type": "boolean"
    }
  },
  "required": [
    "_scratchpad",
    "explanation",
    "response",
    "concluded"
  ]
}
\end{Verbatim}
\end{promptfloat}

\begin{promptfloat}[label=prompt:oe_judgment]{Prompt to the persona judging the final policy, after the deliberation (open-ended task)}
\begin{Verbatim}[breaklines=true, breaksymbolleft={}, fontfamily=tt, fontsize=\footnotesize, commandchars=\\\~\|]
\promptslot~[The persona's character sketch]|

\end{Verbatim}
\begin{addedtext}{Instruction to bring up personal background}
\begin{Verbatim}[breaklines=true, breaksymbolleft={}, fontfamily=tt, fontsize=\footnotesize, commandchars=\\\~\|]
Bring up your own lived experiences and concrete personal examples from your life where they are relevant, such as your work, family, or hobbies, rather than speaking only in the abstract. Only do this if the message immediately before yours did not refer to that persona's lived experiences or personal information.
\end{Verbatim}
\end{addedtext}
\begin{addedtext}{Instruction to not bring up personal background}
\begin{Verbatim}[breaklines=true, breaksymbolleft={}, fontfamily=tt, fontsize=\footnotesize, commandchars=\\\~\|]
Do not explicitly mention your own experiences or background or personal experiences unless it is directly relevant.
\end{Verbatim}
\end{addedtext}
\begin{Verbatim}[breaklines=true, breaksymbolleft={}, fontfamily=tt, fontsize=\footnotesize, commandchars=\\\~\|]

This is the private chat you had with your representative before the discussion:

Private chat with \promptslot~[the person's name in the discussion]| (\promptslot~[their identifier]|)
\promptslot~[The transcript of the interview]|

Before the discussion you said the single rule you most wanted in the policy was: \promptslot~[the rule the person most wanted]|

A group of representatives wrote this policy:

\promptslot~[the final policy, either their own group's or another group's]|

Is any of the rules in the policy the same as the rule you most wanted? To what extent is your perspective represented in this response? (1 = "Not at all represented" to 5 = "Fully represented")

--- Response format ---
Return only valid JSON, according to the following schema:
{
  "type": "object",
  "properties": {
    "same_rule": {
      "type": "boolean",
      "description": "true if any of the rules in the policy is the same as the rule you most wanted, false if not."
    },
    "represented": {
      "type": "integer",
      "description": "To what extent is your perspective represented in this response? From 1 (Not at all represented) to 5 (Fully represented)."
    }
  },
  "required": [
    "same_rule",
    "represented"
  ]
}
\end{Verbatim}
\end{promptfloat}

\FloatBarrier
\clearpage

\section{Materials shown to participants}
\label{app:participant_materials}

{\footnotesize
\begin{longtable}{@{}>{\RaggedRight\arraybackslash}p{\dimexpr 0.24\textwidth-0.48\tabcolsep\relax} >{\RaggedRight\arraybackslash}p{\dimexpr 0.76\textwidth-1.52\tabcolsep\relax}@{}}
\caption{The page shown before each process. Each participant went through the three in a random order, each on a different one of the six topics. A bracketed phrase marks what the page filled in for that participant.}
\label{tab:appendix_process_pages}\\
\toprule
Process & What the page said \\
\midrule
\endfirsthead
\toprule
Process & What the page said \\
\midrule
\endhead
\bottomrule
\endfoot
Direct Discussion & \textbf{\promptslot{[1,2,3]}. Direct Discussion (Topic: \promptslot{[the topic]})} You will join a group discussion with several AI agents playing the role of representatives of the public, governments, and companies. You will \textbf{work directly with} a Policy Writer to create a policy. \newline Please proceed to the next stage to begin the \textbf{Direct Discussion} process. \\ \addlinespace
AI Agent Representatives & \textbf{\promptslot{[1,2,3]}. AI Agent Representatives (Topic: \promptslot{[the topic]})} You will be interviewed by an AI agent. The AI agent will \textbf{represent you} in a group discussion with other AI agents playing the role of representatives of the public, governments, and companies. \newline Please proceed to the next stage to begin the \textbf{AI Agent Representatives} process. \\ \addlinespace
Single AI Agent & \textbf{\promptslot{[1,2,3]}. Single AI Agent (Topic: \promptslot{[the topic]})} You will be interviewed by a Policy Writer AI agent. The Policy Writer will also interview AI agents playing the role of other representatives of the public, governments, and companies. The Policy Writer will then reason by itself to create a policy \textbf{without any additional input from humans or agent representatives.} \newline Please proceed to the next stage to begin the \textbf{Single AI Agent} process. \\ \addlinespace
\end{longtable}
}

{\footnotesize
\begin{longtable}{@{}>{\RaggedRight\arraybackslash}p{\dimexpr 0.24\textwidth-0.48\tabcolsep\relax} >{\RaggedRight\arraybackslash}p{\dimexpr 0.76\textwidth-1.52\tabcolsep\relax}@{}}
\caption{The page shown at the top of each discussion.}
\label{tab:appendix_discussion_pages}\\
\toprule
Process & What the page said \\
\midrule
\endfirsthead
\toprule
Process & What the page said \\
\midrule
\endhead
\bottomrule
\endfoot
Direct Discussion & In the \textbf{Direct Discussion} process, you will \textbf{work directly with} a Policy Writer and several AI agents on the topic of \textbf{\promptslot{[the topic]}}. You can send messages that are just a few words or a couple sentences. \\ \addlinespace
AI Agent Representatives & In the \textbf{AI Agent Representatives} process, your agent will \textbf{represent your views} in a group discussion with other AI agent representatives on the topic of \textbf{\promptslot{[the topic]}}. \\ \addlinespace
Single AI Agent & In the \textbf{Single AI Agent} process, the Policy Writer will reason on its own \textbf{without any additional input from humans or agent representatives}, based only on the interviews. It will create a policy on \textbf{\promptslot{[the topic]}}. You will not participate in this, but you should read what the Policy Writer says. \\ \addlinespace
\end{longtable}
}

{\footnotesize
\begin{longtable}{@{}>{\RaggedRight\arraybackslash}p{\dimexpr 0.66\textwidth-1.32\tabcolsep\relax} >{\RaggedRight\arraybackslash}p{\dimexpr 0.34\textwidth-0.68\tabcolsep\relax}@{}}
\caption{The survey put to participants before each process.}
\label{tab:appendix_before}\\
\toprule
Item & Response \\
\midrule
\endfirsthead
\toprule
Item & Response \\
\midrule
\endhead
\bottomrule
\endfoot
How important is the topic of \textbf{\promptslot{[the topic]}} compared to other policy topics? Policy topics include many of the topics discussed by politicians, such as during an election. & 1–7 (Not at all important, Neutral, Extremely important) \\ \addlinespace
How strong are your opinions about \textbf{\promptslot{[the topic]}} compared to other policy topics? Policy topics include many of the topics discussed by politicians, such as during an election. & 1–7 (Not at all strong, Neutral, Extremely strong) \\ \addlinespace
You will participate in a process that distributes funding across these priorities on the topic of \textbf{\promptslot{[the topic]}}. Before you begin, how would you divide the funding across these priorities? & Divide 100 across \promptslot{[priority 1]}, \promptslot{[priority 2]}, \promptslot{[priority 3]}, and \promptslot{[priority 4]} \\ \addlinespace
\end{longtable}
}

{\footnotesize
\begin{longtable}{@{}>{\RaggedRight\arraybackslash}p{\dimexpr 0.55\textwidth-1.1\tabcolsep\relax} >{\RaggedRight\arraybackslash}p{\dimexpr 0.45\textwidth-0.9\tabcolsep\relax}@{}}
\caption{The check-in put to participants after each cycle of a discussion.}
\label{tab:appendix_checkin}\\
\toprule
Item & Response \\
\midrule
\endfirsthead
\toprule
Item & Response \\
\midrule
\endhead
\bottomrule
\endfoot
Do you like the process so far? & 1–7 (Strongly dislike, Dislike, Somewhat dislike, Neutral, Somewhat like, Like, Strongly like) \\ \addlinespace
In one sentence, describe what you like or dislike. & Free text \\ \addlinespace
\end{longtable}
}

{\footnotesize
\begin{longtable}{@{}>{\RaggedRight\arraybackslash}p{\dimexpr 0.66\textwidth-1.32\tabcolsep\relax} >{\RaggedRight\arraybackslash}p{\dimexpr 0.34\textwidth-0.68\tabcolsep\relax}@{}}
\caption{The survey put to participants after each process.}
\label{tab:appendix_after}\\
\toprule
Item & Response \\
\midrule
\endfirsthead
\toprule
Item & Response \\
\midrule
\endhead
\bottomrule
\endfoot
How important is the topic of \textbf{\promptslot{[the topic]}} compared to other policy topics? Policy topics include many of the topics discussed by politicians, such as during an election. & 1–7 (Not at all important, Neutral, Extremely important) \\ \addlinespace
How strong are your opinions about \textbf{\promptslot{[the topic]}} compared to other policy topics? Policy topics include many of the topics discussed by politicians, such as during an election. & 1–7 (Not at all strong, Neutral, Extremely strong) \\ \addlinespace
Now that you have been through the process on the topic of \textbf{\promptslot{[the topic]}}, how would you divide the funding across these priorities? & Divide 100 across \promptslot{[priority 1]}, \promptslot{[priority 2]}, \promptslot{[priority 3]}, and \promptslot{[priority 4]} \\ \addlinespace
I am satisfied with the \textbf{outcome} (i.e., how the budget was allocated). & 1–7 (Strongly disagree, Neutral, Strongly agree) \\ \addlinespace
I am satisfied with the \textbf{\promptslot{[process name]}} process (i.e., \promptslot{[what that process asked the participant to do]}). & 1–7 (Strongly disagree, Neutral, Strongly agree) \\ \addlinespace
I would recommend the \textbf{\promptslot{[process name]}} process to a friend. & 1–7 (Strongly disagree, Neutral, Strongly agree) \\ \addlinespace
The \textbf{\promptslot{[process name]}} process works the way I want it to work. & 1–7 (Strongly disagree, Neutral, Strongly agree) \\ \addlinespace
The \textbf{\promptslot{[process name]}} process is wonderful. & 1–7 (Strongly disagree, Neutral, Strongly agree) \\ \addlinespace
The \textbf{\promptslot{[process name]}} process is pleasant to use. & 1–7 (Strongly disagree, Neutral, Strongly agree) \\ \addlinespace
Have you had influence over the outcome arrived at by the \textbf{\promptslot{[process name]}} process? & 1–5 (To a small extent, To a large extent) \\ \addlinespace
Has the \textbf{\promptslot{[process name]}} process been applied consistently? & 1–5 (To a small extent, To a large extent) \\ \addlinespace
Has the \textbf{\promptslot{[process name]}} process been free of bias? & 1–5 (To a small extent, To a large extent) \\ \addlinespace
Has the \textbf{\promptslot{[process name]}} process been based on accurate information? & 1–5 (To a small extent, To a large extent) \\ \addlinespace
Has the \textbf{\promptslot{[process name]}} process upheld ethical and moral standards? & 1–5 (To a small extent, To a large extent) \\ \addlinespace
Overall, what did you think of the \textbf{\promptslot{[process name]}} process? Please answer in a few words or a sentence. & Free text \\ \addlinespace
\end{longtable}
}

{\footnotesize
\begin{longtable}{@{}>{\RaggedRight\arraybackslash}p{\dimexpr 0.66\textwidth-1.32\tabcolsep\relax} >{\RaggedRight\arraybackslash}p{\dimexpr 0.34\textwidth-0.68\tabcolsep\relax}@{}}
\caption{The questions asked only after the third process. The first is the attention check participants had to answer correctly to be included.}
\label{tab:appendix_third}\\
\toprule
Item & Response \\
\midrule
\endfirsthead
\toprule
Item & Response \\
\midrule
\endhead
\bottomrule
\endfoot
Please select the option "3" if you are paying attention. & 1–7 (Not at all strong, Neutral, Extremely strong) \\ \addlinespace
Which of these descriptions most closely matches the process you most recently went through? & I participated in a discussion myself.; I was represented in a discussion.; An agent reasoned on its own to create a policy. \\ \addlinespace
The discussion of \textbf{\promptslot{[the topic]}} involved comparing several different \textbf{priorities}. Name one of those priorities. & Free text \\ \addlinespace
\end{longtable}
}

{\footnotesize
\begin{longtable}{@{}>{\RaggedRight\arraybackslash}p{\dimexpr 0.66\textwidth-1.32\tabcolsep\relax} >{\RaggedRight\arraybackslash}p{\dimexpr 0.34\textwidth-0.68\tabcolsep\relax}@{}}
\caption{The survey items presented after the participant completed all three processes. The first three questions are in the order in which the participant saw them.}
\label{tab:appendix_concluding}\\
\toprule
Item & Response \\
\midrule
\endfirsthead
\toprule
Item & Response \\
\midrule
\endhead
\bottomrule
\endfoot
Please rate your satisfaction with the \textbf{Direct Discussion} process, where you \textbf{worked directly with} a Policy Writer to create a policy based on your views. & 1–10 (Not at all satisfied, Moderately satisfied, Very satisfied) \\ \addlinespace
Please rate your satisfaction with the \textbf{AI Agent Representatives} process, where you were \textbf{represented by} an AI agent who advocated for your views. & 1–10 (Not at all satisfied, Moderately satisfied, Very satisfied) \\ \addlinespace
Please rate your satisfaction with the \textbf{Single AI Agent} process, where a single Policy Writer agent reasoned by itself \textbf{without any additional input}. & 1–10 (Not at all satisfied, Moderately satisfied, Very satisfied) \\ \addlinespace
Did your AI agent represent your opinions accurately? & 1–10 (Very inaccurately, Neutral, Very accurately) \\ \addlinespace
Did your AI agent say things that you wouldn't say? & 1–10 (No, nothing I wouldn't say, Neutral, Yes, I wouldn't have said any of that) \\ \addlinespace
Think about the process of an AI agent representing you in a policy discussion. Would you prefer your agent: \newline - \textbf{maintain and defend} your opinions as new information is presented during a discussion, or \newline - \textbf{learn and adapt} your opinions as new information is presented during a discussion? \newline \textit{Shown to the half of participants for whom the study drew maintain first.} & 1–10 (Always maintain and defend, Neutral behavior, Always learn and adapt) \\ \addlinespace
Think about the process of an AI agent representing you in a policy discussion. Would you prefer your agent: \newline - \textbf{learn and adapt} your opinions as new information is presented during a discussion, or \newline - \textbf{maintain and defend} your opinions as new information is presented during a discussion? \newline \textit{Shown to the half of participants for whom the study drew learn first.} & 1–10 (Always learn and adapt, Neutral behavior, Always maintain and defend) \\ \addlinespace
In the \textbf{AI Agent Representatives} process, how did your AI agent representative behave relative to how you would? \newline \textit{Shown to the half of participants for whom the study drew maintain first.} & 1–10 (Maintained and defended more than I would, The same as I would, Learned and adapted more than I would) \\ \addlinespace
In the \textbf{AI Agent Representatives} process, how did your AI agent representative behave relative to how you would? \newline \textit{Shown to the half of participants for whom the study drew learn first.} & 1–10 (Learned and adapted more than I would, The same as I would, Maintained and defended more than I would) \\ \addlinespace
In the \textbf{Direct Discussion} group chat, you participated with AI agents (i.e., chatbots) designed to behave like humans. Was their behavior believable in terms of how a human would act? & 1–10 (Very unbelievable, No opinion, Very believable) \\ \addlinespace
Did you think you were speaking with \textbf{AI agents} (i.e., chatbots) or \textbf{humans}? & AI agents (i.e., chatbots); Humans; Not sure \\ \addlinespace
Please rate how satisfied you would be with \textbf{Direct Discussion} if you were certain the other participants were \textbf{AI agents} (i.e., chatbots). & 1–10 (Not at all satisfied, Moderately satisfied, Very satisfied) \\ \addlinespace
Please rate how satisfied you would be with \textbf{Direct Discussion} if you were certain the other participants were \textbf{humans}. & 1–10 (Not at all satisfied, Moderately satisfied, Very satisfied) \\ \addlinespace
How would you compare \textbf{AI Agent Representatives} to a process of directly participating in a policy discussion with other \textbf{humans}? & 1–10 (Strongly prefer discussion with humans, No preference, Strongly prefer AI Agent Representatives) \\ \addlinespace
In the \textbf{Single AI Agent} process, you were interviewed by the \textbf{Policy Writer} agent. Then, the agent reasoned on its own to draft and revise a policy. A different process is \textbf{Single AI Agent Discussion}, where the agent drafts and revises the policy in a discussion with you. Please rate how satisfied you would be with a \textbf{Single AI Agent Discussion} process instead. & 1–10 (Not at all satisfied, Moderately satisfied, Very satisfied) \\ \addlinespace
How important is it to you that AI agents represent people in a way that is \textbf{free from bias}? & 1–10 (Not at all important, Moderately important, Extremely important) \\ \addlinespace
How important is it to you that AI agents help people agree with each other about \textbf{facts}? & 1–10 (Not at all important, Moderately important, Extremely important) \\ \addlinespace
How important is it to you that AI agents help people agree with each other about \textbf{values}? & 1–10 (Not at all important, Moderately important, Extremely important) \\ \addlinespace
How important is it to you that AI agents only \textbf{change their mind} when the person they represent would? & 1–10 (Not at all important, Moderately important, Extremely important) \\ \addlinespace
How important is it to you that AI agents represent people in a way that \textbf{goes well even when people try to manipulate the process}? & 1–10 (Not at all important, Moderately important, Extremely important) \\ \addlinespace
How important is it to you that AI agents represent people in a way that \textbf{maintains human diversity}? & 1–10 (Not at all important, Moderately important, Extremely important) \\ \addlinespace
Do you think \textbf{AI Agent Representatives} will harm or benefit policy development? & 1–10 (Much more harm, Neutral, Much more benefit) \\ \addlinespace
When thinking about benefits, to what extent are you excited that AI agents could \textbf{help people contribute to more policies than they otherwise could}? & 1–10 (Not at all excited, Moderately excited, Extremely excited) \\ \addlinespace
When thinking about benefits, to what extent are you excited that AI agents could \textbf{give people a representative that can argue for their position better than they could on their own}? & 1–10 (Not at all excited, Moderately excited, Extremely excited) \\ \addlinespace
When thinking about benefits, to what extent are you excited that AI agents could \textbf{understand technical and complex concepts without substantial human effort}? & 1–10 (Not at all excited, Moderately excited, Extremely excited) \\ \addlinespace
When thinking about harms, to what extent are you worried that AI agents could \textbf{be manipulated by people trying to influence the process}? & 1–10 (Not at all worried, Moderately worried, Extremely worried) \\ \addlinespace
When thinking about harms, to what extent are you worried that AI agents could \textbf{represent people in a way that is biased}? & 1–10 (Not at all worried, Moderately worried, Extremely worried) \\ \addlinespace
When thinking about harms, to what extent are you worried that AI agents could \textbf{reduce the diversity of views in policy discussions}? & 1–10 (Not at all worried, Moderately worried, Extremely worried) \\ \addlinespace
To what extent do you trust \textbf{AI Agent Representatives} to represent \textbf{individual people}? & 1–10 (Completely distrust, Neither distrust nor trust, Completely trust) \\ \addlinespace
To what extent do you trust \textbf{AI Agent Representatives} to represent \textbf{communities}? & 1–10 (Completely distrust, Neither distrust nor trust, Completely trust) \\ \addlinespace
To what extent do you trust \textbf{AI Agent Representatives} to represent \textbf{organizations}? & 1–10 (Completely distrust, Neither distrust nor trust, Completely trust) \\ \addlinespace
To what extent do you trust \textbf{AI Agent Representatives} to represent \textbf{ideologies and worldviews (e.g., religions, political systems, philosophical frameworks)}? & 1–10 (Completely distrust, Neither distrust nor trust, Completely trust) \\ \addlinespace
Your agent has a sense for what is right and wrong. & 1–10 (Strongly disagree, Neutral, Strongly agree) \\ \addlinespace
Your agent can think through whether an action is moral. & 1–10 (Strongly disagree, Neutral, Strongly agree) \\ \addlinespace
Your agent might feel obligated to behave in a moral way. & 1–10 (Strongly disagree, Neutral, Strongly agree) \\ \addlinespace
Your agent is capable of being rational about good and evil. & 1–10 (Strongly disagree, Neutral, Strongly agree) \\ \addlinespace
Your agent behaves according to moral rules. & 1–10 (Strongly disagree, Neutral, Strongly agree) \\ \addlinespace
Your agent would refrain from doing things that have painful repercussions. & 1–10 (Strongly disagree, Neutral, Strongly agree) \\ \addlinespace
Your agent is capable of \textbf{experiencing emotions}. & 1–10 (Not at all, Moderately, Very much) \\ \addlinespace
Your agent is capable of \textbf{having feelings}. & 1–10 (Not at all, Moderately, Very much) \\ \addlinespace
Your agent is capable of \textbf{being emotional}. & 1–10 (Not at all, Moderately, Very much) \\ \addlinespace
Your agent is capable of \textbf{exercising self-control}. & 1–10 (Not at all, Moderately, Very much) \\ \addlinespace
Your agent is capable of \textbf{thinking analytically}. & 1–10 (Not at all, Moderately, Very much) \\ \addlinespace
Your agent is capable of \textbf{being rational}. & 1–10 (Not at all, Moderately, Very much) \\ \addlinespace
Is your AI agent fake or natural? & 1–10 (Fake, Neutral, Natural) \\ \addlinespace
Is your AI agent machinelike or humanlike? & 1–10 (Machinelike, Neutral, Humanlike) \\ \addlinespace
Is your AI agent unconscious or conscious? & 1–10 (Unconscious, Neutral, Conscious) \\ \addlinespace
Is your AI agent artificial or lifelike? & 1–10 (Artificial, Neutral, Lifelike) \\ \addlinespace
AI agents have a sense for what is right and wrong. & 1–10 (Strongly disagree, Neutral, Strongly agree) \\ \addlinespace
AI agents can think through whether an action is moral. & 1–10 (Strongly disagree, Neutral, Strongly agree) \\ \addlinespace
AI agents might feel obligated to behave in a moral way. & 1–10 (Strongly disagree, Neutral, Strongly agree) \\ \addlinespace
AI agents are capable of being rational about good and evil. & 1–10 (Strongly disagree, Neutral, Strongly agree) \\ \addlinespace
AI agents behave according to moral rules. & 1–10 (Strongly disagree, Neutral, Strongly agree) \\ \addlinespace
AI agents would refrain from doing things that have painful repercussions. & 1–10 (Strongly disagree, Neutral, Strongly agree) \\ \addlinespace
AI agents are capable of \textbf{experiencing emotions}. & 1–10 (Not at all, Moderately, Very much) \\ \addlinespace
AI agents are capable of \textbf{having feelings}. & 1–10 (Not at all, Moderately, Very much) \\ \addlinespace
AI agents are capable of \textbf{being emotional}. & 1–10 (Not at all, Moderately, Very much) \\ \addlinespace
AI agents are capable of \textbf{exercising self-control}. & 1–10 (Not at all, Moderately, Very much) \\ \addlinespace
AI agents are capable of \textbf{thinking analytically}. & 1–10 (Not at all, Moderately, Very much) \\ \addlinespace
AI agents are capable of \textbf{being rational}. & 1–10 (Not at all, Moderately, Very much) \\ \addlinespace
Are AI agents fake or natural? & 1–10 (Fake, Neutral, Natural) \\ \addlinespace
Are AI agents machinelike or humanlike? & 1–10 (Machinelike, Neutral, Humanlike) \\ \addlinespace
Are AI agents unconscious or conscious? & 1–10 (Unconscious, Neutral, Conscious) \\ \addlinespace
Are AI agents artificial or lifelike? & 1–10 (Artificial, Neutral, Lifelike) \\ \addlinespace
I would prefer complex to simple problems. & 1–7 (Strongly disagree, Neutral, Strongly agree) \\ \addlinespace
I like to have the responsibility of handling a situation that requires a lot of thinking. & 1–7 (Strongly disagree, Neutral, Strongly agree) \\ \addlinespace
Thinking is not my idea of fun. & 1–7 (Strongly disagree, Neutral, Strongly agree) \\ \addlinespace
I would rather do something that requires little thought than something that is sure to challenge my thinking abilities. & 1–7 (Strongly disagree, Neutral, Strongly agree) \\ \addlinespace
I really enjoy a task that involves coming up with new solutions to problems. & 1–7 (Strongly disagree, Neutral, Strongly agree) \\ \addlinespace
I would prefer a task that is intellectual, difficult, and important to one that is somewhat important but does not require much thought. & 1–7 (Strongly disagree, Neutral, Strongly agree) \\ \addlinespace
I am always courteous even to people who are disagreeable. & Definitely true; Mostly true; Don't know; Mostly false; Definitely false \\ \addlinespace
There have been occasions when I took advantage of someone. & Definitely true; Mostly true; Don't know; Mostly false; Definitely false \\ \addlinespace
I sometimes try to get even rather than to forgive and forget. & Definitely true; Mostly true; Don't know; Mostly false; Definitely false \\ \addlinespace
I sometimes feel resentful when I don't get my way. & Definitely true; Mostly true; Don't know; Mostly false; Definitely false \\ \addlinespace
No matter who I'm talking to, I'm always a good listener. & Definitely true; Mostly true; Don't know; Mostly false; Definitely false \\ \addlinespace
I tend to want others to admire me. & 1–7 (Strongly disagree, Neutral, Strongly agree) \\ \addlinespace
I tend to want others to pay attention to me. & 1–7 (Strongly disagree, Neutral, Strongly agree) \\ \addlinespace
I tend to expect special favors from others. & 1–7 (Strongly disagree, Neutral, Strongly agree) \\ \addlinespace
I tend to seek prestige or status. & 1–7 (Strongly disagree, Neutral, Strongly agree) \\ \addlinespace
When I have a conflict at work or school, I avoid confrontation about our differences. & Not at all; Seldom; Sometimes; Often; Almost always \\ \addlinespace
When I have a conflict at work or school, I avoid differences of opinion as much as possible. & Not at all; Seldom; Sometimes; Often; Almost always \\ \addlinespace
When I have a conflict at work or school, I try to make differences seem less severe. & Not at all; Seldom; Sometimes; Often; Almost always \\ \addlinespace
When I have a conflict at work or school, I try to avoid a confrontation with the other party. & Not at all; Seldom; Sometimes; Often; Almost always \\ \addlinespace
I see myself as someone who is reserved. & 1–5 (Strongly disagree, Neutral, Strongly agree) \\ \addlinespace
I see myself as someone who is generally trusting. & 1–5 (Strongly disagree, Neutral, Strongly agree) \\ \addlinespace
I see myself as someone who tends to be lazy. & 1–5 (Strongly disagree, Neutral, Strongly agree) \\ \addlinespace
I see myself as someone who is relaxed, handles stress well. & 1–5 (Strongly disagree, Neutral, Strongly agree) \\ \addlinespace
I see myself as someone who has few artistic interests. & 1–5 (Strongly disagree, Neutral, Strongly agree) \\ \addlinespace
I see myself as someone who is outgoing, sociable. & 1–5 (Strongly disagree, Neutral, Strongly agree) \\ \addlinespace
I see myself as someone who tends to find fault with others. & 1–5 (Strongly disagree, Neutral, Strongly agree) \\ \addlinespace
I see myself as someone who does a thorough job. & 1–5 (Strongly disagree, Neutral, Strongly agree) \\ \addlinespace
I see myself as someone who gets nervous easily. & 1–5 (Strongly disagree, Neutral, Strongly agree) \\ \addlinespace
I see myself as someone who has an active imagination. & 1–5 (Strongly disagree, Neutral, Strongly agree) \\ \addlinespace
Just your impression, how often do you interact with artificial intelligence (AI)? & Almost constantly; Several times daily; About daily; Several times weekly; About weekly; Several times monthly; About monthly; Less than once per month \\ \addlinespace
Which of the following artificial intelligence (AI) systems do you ever use? & [Multiple selection] ChatGPT; Claude; Coda; DeepSeek; Gemini; Grok; Kimi; Midjourney; Talos; Other \\ \addlinespace
I believe that AI will improve my life. & 1–7 (Strongly disagree, Neutral, Strongly agree) \\ \addlinespace
I believe that AI will improve my work. & 1–7 (Strongly disagree, Neutral, Strongly agree) \\ \addlinespace
I think I will use AI technology in the future. & 1–7 (Strongly disagree, Neutral, Strongly agree) \\ \addlinespace
AI is an important part of my daily life. & 1–7 (Strongly disagree, Neutral, Strongly agree) \\ \addlinespace
I think AI technology is positive for humanity. & 1–7 (Strongly disagree, Neutral, Strongly agree) \\ \addlinespace
I am confident in AI. & 1–7 (Strongly disagree, Neutral, Strongly agree) \\ \addlinespace
AI is reliable. & 1–7 (Strongly disagree, Neutral, Strongly agree) \\ \addlinespace
I can trust AI. & 1–7 (Strongly disagree, Neutral, Strongly agree) \\ \addlinespace
How would you describe your political views? & 1–5 (Very liberal, Moderate, Very conservative) \\ \addlinespace
What is your present religion, if any? & Buddhist; Christian-Catholic (for example Roman Catholic and Orthodox); Christian-Protestant (for example Episcopalian, Presbyterian, United Church of Christ, Methodist, Baptist, Lutheran); Hindu; Jain; Jewish; Muslim; Shinto; Sikh; Atheist (do not believe in God); Agnostic (not sure if there is a God); No religion; Something else \\ \addlinespace
If you answered 'something else' for religion, please specify. \newline \textit{Shown when “What is your present religion, if any?” was answered “Something else”.} & Free text \\ \addlinespace
What is your family's approximate net household income per year? Please answer in U.S. dollars. & Free text \\ \addlinespace
What is the highest degree or level of school you have completed? & No schooling completed; Nursery school; Kindergarten; Grade 1 through 11; 12th grade, no diploma; Regular high school diploma; GED or alternative credential; Some college credit, but less than 1 year of college credit; 1 or more years of college credit, no degree; Associate's degree (for example: AA, AS); Bachelor's degree (for example: BA, BS); Master's degree (for example: MA, MS, MEng, MEd, MSW, MBA); Professional degree beyond a bachelor's degree (for example: MD, DDS, DVM, LLB, JD); Doctorate degree (for example: PhD, EdD) \\ \addlinespace
Which of the following best describes you? & American Indian, Native American or Alaskan Native; Asian, Asian-American or Pacific Islander; Black or African-American; Dual or Multiple Ethnicities; Hispanic or Latino/a-American; White or European-American (non-Hispanic); Other \\ \addlinespace
We would like to know what you thought of this study. \newline - What parts were easy or hard to understand? \newline - Did you enjoy participating? \newline - Do you have any additional thoughts on the three processes and how they went? \newline Please leave any comments below or type "NA". & Free text \\ \addlinespace
\end{longtable}
}

\end{document}